*Journal of Software Engineering Research and Development, 2021, 9:5,* doi: 10.5753/jserd.2021.1096This work is licensed under a Creative Commons Attribution 4.0 International License.# Linking Use Cases and Associated Requirements: A Replicated Eye Tracking Study on the Impact of Linking Variants on Reading Behavior

**Oliver Karras**  [ Leibniz Information Centre for Science and Technology | *oliver.karras@tib.eu* ]
**Alexandra Risch** [ Leibniz Universität Hannover | *alexandra.risch@se.uni-hannover.de* ]
**Jil Klünder**  [ Leibniz Universität Hannover | *jil.kluender@inf.uni-hannover.de* ]**Abstract**

A wide variety of use case templates supports different variants to link a use case with its associated requirements. Regardless of the linking, a reader must process the related information simultaneously to understand them. Linking variants are intended to cause a specific reading behavior in which a reader interrelates a use case and its associated requirements. Due to the effort to create and maintain links, we investigated the impact of different linking variants on the reading behavior in terms of visual effort and the intended way of interrelating both artifacts. We designed an eye tracking study about reading a use case and requirements. We conducted the study twice each with 15 subjects as a baseline experiment and as a repetition. The results of the baseline experiment, its repetition, and their joint analysis are consistent. All investigated linking variants cause comparable visual effort. In all cases, reading the single artifacts one after the other is the most frequently occurring behavior. Only links embedded in the fields of a use case description significantly increase the readers' efforts to interrelate both artifacts. None of the investigated linking variants impedes reading a use case and requirements. However, only the most detailed linking variant causes readers to process related information simultaneously.

**Keywords:** *Linking, use case, requirement, reading behavior, eye tracking, visual effort, attention switch, cognitive load*## 1 Introduction: Linking Use Cases

Requirements specifications contain several artifact types such as descriptions of interactions or system functions. Gross and Doerr (2012a,b), as well as Ahrens et al. (2016), found that different roles in software development, such as architects and developers, focus on these descriptions when reading a specification. Common notations to represent interactions and functionalities are *fully dressed use cases* and natural language requirements (Fricker et al., 2015).

*Fully dressed use case* templates prescribe a number of elements, such as trigger, precondition, basic sequence, or extensions. Anda et al. (2001) compared the use of different guidelines for writing use cases, including use case templates. They showed that templates are significantly more useful to write high-quality and easy-to-understand use cases than guidelines (Anda et al., 2001).

The original use case template by Jacobson (1993) consisted of natural language descriptions embedded in a table. Coleman (1998) argued that this template was incomplete, and the absence of a UML standard resulted in a wide variety of use case templates since users were likely to invent their own format. He proposed a UML compatible template which was more complete and less ambiguous (Coleman, 1998). Thus, this template should be easier to use. Coleman (1998) included a field for non-functional requirements since use cases often involve additional information that does not fit typical use case fields (Wiegers and Beatty, 2013). However, the idea of attaching associated requirements to a use case stimulated the proliferation of use case variants (Tiwari and Gupta, 2015). Researchers and practitioners invented further formats to add any associated information that supports a reader's understanding or is valuable for a project (Cockburn, 2001; Kruchten, 2004). However, there is no consensus about how to relate a use case to its associated requirements (Tiwari and Gupta, 2015).

The diverse use case templates share typical fields such as title and basic sequence (Anda et al., 2001; Tiwari and Gupta, 2015). Nevertheless, these templates also include various linking variants referring to different associated information such as functional (Alexander and Neil, 2005), non-functional (Coleman, 1998), or special requirements (Kruchten, 2004). Based on the literature, we identified three widely used linking variants. Besides *no linking* (Jacobson, 1993; Schneider and Winters, 1998; Leite et al., 2000; Kettenis, 2007; Kulak and Guiney, 2012), templates mainly include an *additional field* to list all associated requirements (Coleman, 1998; Insfrán et al., 2002; Liu et al., 2003; Kruchten, 2004; Zhou et al., 2014), or when using a requirements management tool *integrated links* in typical fields refer to the associated requirements (Wiegers, 1999; Cockburn, 2001; Wiegers and Beatty, 2013).

The variants *additional field* and *integrated links* permit traceability from a use case to its associated requirements. Traceability is one of the key characteristics of excellent specifications since it provides several benefits such as change impact analysis or supported document maintenance (Wiegers and Beatty, 2013). The main purpose of both linking variants is to highlight the interrelationships between a use case and its associated requirements. Created links are intended to cause a specific reading behavior. A reader should follow the links to interrelate both artifacts in order to process them. Hereafter, we refer to this specific reading behavior as *the intended way of interrelating both artifacts*.

Regardless of the particular linking variant, creating and maintaining links is a challenging task that increases the development costs by the effort to accumulate and manage the traceability information (Wiegers and Beatty, 2013). If this information becomes obsolete, readers waste time due to following wrong links. Therefore, defining and maintaining links is "*an investment that increases the chances of delivering a maintainable product that satisfies all the stated customer requirements*" (Wiegers and Beatty, 2013, p. 358).

According to Wiegers and Beatty (2013) as well as Robertson and Robertson (2012), links are mainly realized by adding the labels of the associated requirements to a use case. These labels consist of unique and persistent identification numbers (Wiegers and Beatty, 2013; Robertson and Robertson, 2012). However, this kind of link is troublesome. Basirati et al. (2015) performed a case study to understand changes in use cases. Their results showed that links based on identification numbers are one specific source of risky, dispersed changes (Basirati et al., 2015). "*These types of numbered references are very hard to maintain and can easily lead to wrong references*" (Basirati et al., 2015, p. 360).

Due to the effort to create and maintain links, we investigate whether the three linking variants (*no linking*, *additional field*, and *integrated links*) have an impact on the reading behavior. We ask the following research question:

> **Research question:**
> How does the linking variant between a use case and its associated requirements influence the reading behavior in terms of visual effort and intended way of interrelating both artifacts?

# 2 Research Approach

In the following, we present the research approach to investigate the aforementioned research question. We explain the details of the research process and its contribution.

## 2.1 Research Process

The research process is based on the *multiple-replication types approach* by Gómez et al. (2014). In this way, we follow a systematic approach to study the phenomenon of the impact of linking variants on reading behavior in software engineering experimentally.

In general, we developed an eye tracking study for a between-subjects experiment to compare the three previously mentioned linking variants. Each linking variant was applied to the same use case and requirements so that all used materials differ only in the respective linking variant. Thus, we could investigate whether the different linking variants have an impact on how the subjects read the same materials to process them. We used eye tracking since this technology enables to capture and analyze the individual reading behavior of the subjects in detail. Based on the collected eye tracking data, we investigated the visual effort and the reading behavior of each subject for the particular linking variant.

We conducted the experiment twice, as a baseline experiment and as a repetition. As a first step, the baseline experiment is the initial implementation of the developed eye tracking study. However, after just one experiment, we do not know whether the observed results are a product of chance. For this reason, the second step towards verifying the results is to repeat the experiment (Gómez et al., 2014). In a repetition, the same experimenters repeat the same experiment at the same site, using the same protocol, with the same operationalization, on a different sample of the same population. The kind of replication of an experiment helps to determine the natural variation of the observed results, i.e., the confidence interval within which the results are observed (Gómez et al., 2014). Thus, a repetition can verify that the results are not a chance product which increases the conclusion validity. However, a single replication of an experiment does not represent the end of the research process. Instead, the repetition is the starting point of a learning process about the experimental conditions which may have an influence on the phenomenon under study and should be controlled (Gómez et al., 2014). Future work requires further replications which are progressively more different from the baseline experiment to verify the results and increase their overall validity in terms of external, internal, construct, and conclusion validity.

## 2.2 Contribution

First, this article is an extended version of our full research paper (Karras et al., 2018) presented at the *22nd Evaluation and Assessment in Software Engineering Conference* (EASE2018). In our full research paper (Karras et al., 2018), we published only the baseline experiment. In this article, we report the baseline experiment (Karras et al., 2018), its repetition, the comparison of the results of both experiments, and a joint analysis of both experiments.

In addition to the baseline experiment and its findings, we replicated the exact same eye tracking study with 15 new subjects to verify the experimental results and increase their conclusion validity. In particular, we present the results of the repetition, compare them with the results of the baseline experiment, and highlight their consistent and inconsistent findings. The joint analysis of both experiments resolved the only inconsistent finding between the two experiments which initially appeared to be a difference, but is actually within the range of the natural random variation of the results. Based on the findings of both experiments and their joint analysis, we contribute the following insights:

(a) The baseline experiment, the repetition, and the joint analysis show no significant difference in the visual effort between the three linking variants. We measured three metrics for visual effort based on the number of fixations, the duration of fixations, and the duration of fixations and saccades. All three metrics indicate that all variants cause comparable visual effort.

(b) Both experiments and their joint analysis detect that readers mostly read the single artifacts separately and successively. Nevertheless, the results of both experiments show that especially *integrated links* have an impact on the reading behavior in terms of the intended way of interrelating both artifacts. We analyzed the reading behavior based on scan-paths. Our analyses indicate that a reading behavior only includes interrelating both artifacts in the case of visualized links. However, only *integrated links* differ significantly from *no linking* regarding this specific aspect of reading behavior.

This paper is structured as follows: Section 3 presents the background and discusses related work. Section 4 describes our research method. Section 5 reports the results of the baseline experiment, whereas section 6 accounts for the repetition and its results. Section 7 compares the results of both experiments. Section 8 describes the threats to validity. In section 9, we discuss our findings. Section 10 concludes the paper.

# 3 Background and Related Work

## 3.1 Use Case Specification

Use cases are widely accepted and acknowledged for specifying interactions and functionalities (Tiwari and Gupta, 2015). Tiwari and Gupta (2015) conducted a systematic literature review to examine the evolution of the use cases, i.a., regarding their representation. The results of the literature study highlight the inherent variability of use cases templates. While Tiwari and Gupta (2015) found 20 different use case templates in their systematic literature review, we found six more use case templates in our own literature review. In total, we identified 26 different use case templates which evolved over the years. In the following, we present a brief summary of the evolution of these use case templates and their linking variants for associated requirements.

### 3.1.1 Use Case Templates

The 26 use case templates range from informal descriptions in paragraph-style text to formal keyword-oriented templates. This proliferation of use case templates illustrates considerable effort and thoughts on the part of various researchers and practitioners (Tiwari and Gupta, 2015).

Based on a simple paragraph-style format (Jacobson, 1993), the first templates evolved by adding numbered main and alternative event flows to improve the textual specification of use cases (Harwood, 1997; Coleman, 1998; Jaaksi, 1998; Mattingly and Rao, 1998; Schneider and Winters, 1998; Toro et al., 1999; Leite et al., 2000). Around the year 2000, several researchers started to increase the formalism of use case templates by changing the paragraph-style format to a tabular-/column-style format (Fleisch, 1999; Anda et al., 2001; Kujala et al., 2001). In this context, the use of specific keywords and the addition of associated information, i.a., requirements, began to support a reader's understanding and the automated generation of domain models (Cockburn, 2001; Insfrán et al., 2002; Liu et al., 2003; Araujo and Coutinho, 2003; Paech and Kohler, 2004; Kruchten, 2004; Haumer, 2004; Somé, 2006; Kettenis, 2007; El-Attar and Miller, 2009). In recent years, the formalism of use case templates has been further increased by using more constrained forms of modeling languages and breaking up event flows to enable the automated generation of software artifacts such as class or sequence diagrams (Kulak and Guiney, 2012; Tiwari et al., 2012; Yue et al., 2013; Misbhauddin and Alshayeb, 2015; Zhou et al., 2014).

Although use case specifications have been largely applied to various software development activities, a lot of research has been conducted to improve the quality of use case specifications for requirements documentation (Tiwari and Gupta, 2015). Despite these efforts, there are still inconsistencies on how to improve the quality of use case specifications for requirements documentation. One of these inconsistencies is how to deal with associated requirements. While some researchers (Jacobson, 1993; Bittner and Spence, 2003; Öuergaard, 2005) suggest not including requirements in use cases to keep them clear and precise, other researchers (Cockburn, 2001; Kruchten, 2004; Kettenis, 2007) consider requirements an integral part of use cases. In particular, these researchers (Cockburn, 2001; Kruchten, 2004; Kettenis, 2007) emphasize that the linking between a use case specification and its associated requirements is important to support the reader's understanding. However, there is no consensus about how to relate a use case specification to its associated requirements (Tiwari and Gupta, 2015).

### 3.1.2 Linking Use Cases and Requirements

We analyzed the 26 use case templates and the corresponding literature regarding the linking variants between a use case specification and its associated requirements. Based on this analysis we identified three major linking variants: *no linking*, *additional field*, and *integrated links* (see Figure 1).

**No Linking.** This variant does not consider any linking between a use case and its associated requirements (see Figure 1a). Seventeen use case templates do not provide any option to perform linking between the use case specification and its associated requirements (see Table 1). As a consequence, the two artifacts are not related to each other. A reader must interrelate the two artifacts himself to process and understand their interrelationships.

**Additional Field.** This variant provides an additional field to link a use case and its associated requirements (see Figure 1b). We identified nine use case templates that use this linking variant (see Table 1). Besides the initial term of "*non-functional requirements*" by Coleman (1998), sev-eral researchers used various other terms to label an additional field such as *"quality requirements"*, *"usability requirements"*, *"special requirements"*, or *"feature requirement"*. All of these differently labeled fields serve the same purpose of adding the associated requirements to the respective use case specification. As a result, the requirements are presented in the context of the use case. However, the reader still must interrelate the two artifacts himself since the reader only knows which requirements are related to the entire use case, but not to the individual use case fields in detail.

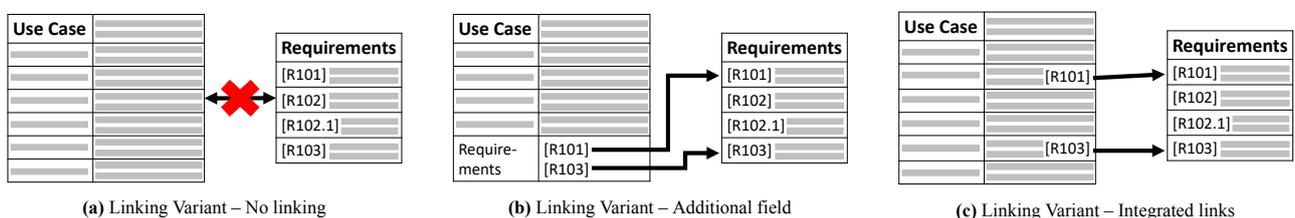

(a) Linking Variant – No linking  (b) Linking Variant – Additional field  (c) Linking Variant – Integrated links

**Figure 1.** Three major linking variants between use cases and associated requirements

Table 1. Three major linking variants between a use case and its associated requirements identified in the literature

| Linking variant | References |
| --- | --- |
| No linking | Jacobson (1993); Harwood (1997); Mattingly and Rao (1998); Schneider and Winters (1998); Toro et al. (1999); Fleisch (1999); Leite et al. (2000); Cockburn (2001); Kujala et al. (2001); Anda et al. (2001); Somé (2006); Kettenis (2007); El-Attar and Miller (2009); Kulak and Guiney (2012); Tiwari et al. (2012); Yue et al. (2013); Misbhauddin and Alshayeb (2015) |
| Additional field | Coleman (1998); Jaaksi (1998); Insfrán et al. (2002); Liu et al. (2003); Araujo and Coutinho (2003); Kruchten (2004); Paech and Kohler (2004); Haumer (2004); Zhou et al. (2014) |
| Integrated links | Wiegers (1999); Cockburn (2001); Wiegers and Beatty (2013) |

**Integrated Links.** This variant includes links to requirements in typical use case fields (see Figure 1c). In the corresponding literature, we found that this linking variant is often realized by using a requirements management tool, such as DOORS by IBM[1]. Wiegers (1999) explains that several requirements management tools support such a detailed linking. According to Cockburn (2001), the use of integrated links is a viable tool for use case specifications since additional data descriptions, such as requirements, are often complex and do not fit neatly into the use case specification. For this reason, the application of integrated links from the use case for example to a page containing the associated requirements is a suitable solution (Cockburn, 2001; Wiegers and Beatty, 2013). As a consequence, the requirements are directly associated with the respective affected fields of the use case. In this way, a reader can simply follow the link to process and understand the interrelationships of the two artifacts.

Regardless of the particular linking variant, a reader must interrelate a use case and its associated requirements to process and understand the related information simultaneously (Sweller et al., 2011). However, the three variants cause a different amount of effort to create and maintain links. Therefore, the question comes up whether all linking variants result in a similar reading behavior.

## 3.2 Reading in Software Engineering

Reading software development artifacts is a primary task in software engineering. *Software reading* is defined as the process by which a reader gets a sufficient understanding of the information encoded in a software development artifact to accomplish a particular task (Shull, 2002). The particular task is related to the purpose of reading such as gaining knowledge, detecting defects, or implementing a design (Zhu, 2016). There are various software reading techniques to support a reader during his consideration of a software development artifact. A software reading technique is a sequence of steps for the individual analysis of a textual document to achieve the understanding needed for a particular task. These steps are a set of instructions that guide a reader on how to read a software development artifact, which areas to focus on, and which problems to look for (Shull, 2002; Zhu, 2016). Common reading techniques are *ad-hoc reading* (Porter et al., 1995), *checklist-based reading* (Fagan, 2002), *defect-based reading* (Porter et al., 1995), *perspective-based reading* (Basili et al., 1996), and *usage-based reading* (Thelin et al., 2001). These reading techniques are used in particular for inspections of software development artifacts, e.g. requirements specifications, which require efficient and effective reading in order to detect defects.

*Ad-hoc reading* implies that a reader considers an artifact based on his own skills and knowledge. There is no specific method to guide a reader to understand the artifact. Therefore, the effectiveness of *ad-hoc reading* strongly depends on the individual reader (Porter et al., 1995). *Checklist-based reading* provides a checklist that a reader uses during the consideration of a software development artifact. Such a checklist often consists of questions whose answers help a reader to focus on specific details of an artifact (Fagan, 2002). *Defect-based reading* has the main idea of distributing different aspects of an artifact to different readers. Thus, each reader only focuses on specific aspects of an artifact while inspecting the whole document (Porter et al., 1995). *Perspective-based reading* utilizes different perspectives such as tester, designer, or user to read a software development artifact. The different perspectives help to reduce overlap of reading by several readers since the different perspectives should focus on specific details related to the respective perspective (Basili et al., 1996). *Usage-based reading* focuses on the utilization of use cases to guide a reader during the consideration of an artifact. Thus, the respective document is read from the user's point of view which should support an efficient understanding in terms of user needs (Thelin et al., 2001). Besides all these reading techniques for single software development artifacts, Travassos et al. (1999) proposed the *traceability-based reading* technique to read two related software development artifacts. *Traceability-based reading* focuses on the inspection of either the consistency of two design documents or the correctness and completeness of one design and one requirements document. This reading technique does not consider the comparison of two linked requirements artifacts such as *fully dressed use cases* and natural language requirements.

---
[1] https://www.ibm.com

There is no final conclusion on which reading technique is the best for inspecting a software development artifact. On the one hand, some empirical investigations showed that *checklist-based reading*, which is one of the most commonly used reading techniques in the software industry (Berling and Thelin, 2004), is not more efficient than *ad-hoc reading*. As for *defect-based reading*, *perspective-based reading*, and *usage-based reading*, these investigations achieved slightly better performance than *checklist-based reading* and *ad-hoc reading* (Porter et al., 1995; Basili et al., 1996; Porter and Votta, 1998; Thelin et al., 2003). On the other hand, Halling et al. (2001) reported the opposite finding that *checklist-based reading* is better than *perspective-based reading*. Several other studies also showed that the different reading techniques have no significant difference (Fusaro et al., 1997; Miller et al., 1998; Sandahl et al., 1998).

Reading is primarily an individual effort (Zhu, 2016). This explains the diverging results among the empirical studies. A reader's individual performance is more dominant than the reading technique itself (Uwano et al., 2006). Hence, instead of focusing on the applied reading technique itself, there is a need for a better understanding of a reader's individual reading behavior. The findings of our baseline experiment show first promising insights that the individual reading behavior can be influenced by the respective linking variant (Karras et al., 2018). However, these insights are only based on one experiment. For this reason, we do not know whether the observed results are a product of chance. Hence, we decided to further investigate the impact of the three different linking variants between a use case and its associated requirements. For this purpose, we followed the systematic approach by Gómez et al. (2014) by repeating the exact same experiment with a different sample of the same population. On the one hand, this procedure helps to verify that the results of the baseline experiment are not a product of chance. On the other hand, the repetition increases the sample size for a joint analysis of the collected data to increase conclusion validity of the findings. In particular, we kept the focus on the reading behavior of a respective reader. The individual way of reading appears in a reader's eye movements. An objective way to capture these eye movements and thus to characterize reading behavior is the use of eye tracking. Eye tracking is a powerful technology that offers multiple objective metrics to assess reading behavior in terms of visual effort and intended way of interrelating both artifacts (Sharafi et al., 2015).

### 3.3 Eye Tracking in Software Engineering

Eye tracking technology became increasingly accepted for empirical studies that analyze reading behavior (Sharafi et al., 2015). Especially, the comparison of alternative designs and layout representations was intensively investigated to understand comprehension tasks and reading behavior.

Yusuf et al. (2007) investigated the impact of several characteristics such as layout, color, and stereotypes on the comprehension of UML class diagrams. Porras and Guéhéneuc (2010) analyzed different UML presentations of design patterns with respect to developers' comprehension. Sharif and Maletic (2010) compared orthogonal and multi-clustered representations of UML class diagrams. They investigated the impact of different layouts on the comprehension of design patterns by developers. Sharafi et al. (2013) focused on the efficiency of graphical vs. textual representations of a specific notation, called TROPOS, for modeling and presenting software requirements. Santos et al. (2016) evaluated the effect of layout guidelines for i* goal models on novice stakeholders' ability to understand and review such models. Karras et al. (2017b) compared the original task board design with three customized ones. They investigated the impact of the different design alternatives on developers' work with and comprehension of a task board. Bednarik and Tukiainen (2006) proposed a visualization technique for source code. They analyzed how developers use the different perspectives of normal and visualized source code interchangeably. Busjahn et al. (2011) investigated the difference between reading source code and natural text in an experiment. Sharafi et al. (2012) compared different representation styles of source code identifiers. They analyzed the recalling of the names of identifiers by considering different strategies deployed by men and women. Binkley et al. (2013) also focused on the impact of identifier styles on the code comprehension of developers. Romero et al. (2002) analyzed the use of different representations by developers while performing debugging tasks and the impact of these representations on the developers' performance. Romero et al. (2003) extended their previous work by characterizing the developers' strategies in debugging tasks based on the level of focus attention, representation use, and reasoning strategy. Hejmady and Narayanan (2012) also studied the effectiveness and role of different representations used during source code debugging. Ali et al. (2012) applied eye tracking to understand how developers verify links for requirements traceability. They identified and ranked the preferred source code entities of developers to define two weighting schemes to recover traceability links.

Sharafi et al. (2015) conducted a systematic literature study on the usage of eye tracking in software engineering. They identified 36 relevant papers (Sharafi et al., 2015). The major three research topics of these 36 papers are *code comprehension* (12 papers), *model comprehension* (10 papers), and *debugging* (9 papers). Thus, the majority of studies used source code and graphical models as objects of investigation. Only two of the 36 papers included English texts as an object of investigation in their study (Sharafi et al., 2015). Although $95\%$ of requirements documents are written in common or structured natural language, e.g. templates or forms (Mich et al., 2004), there are only a few eye tracking studies which address the comprehension of such textual representations so far. Ahrens et al. (2016) conducted an eye tracking study to analyze how requirements specifications are read. They identified similar patterns between paper- and screen-based reading. The results contribute awareness by considering the readers' interests based on how they use a specification. Gross and Doerr (2012a) performed an explorative eye tracking study to investigate software architects' information needs and expectations from a requirements specification. The results provide first insights into the relevance of artifact types and their notational representations. Gross and Doerr (2012b) extended their eye tracking study by analyzing information needs and expectations of usability experts. Based on the findings, they introduced the idea of a view-based requirements specification to fulfill the needs of different roles in software development.

The number of eye tracking studies that focus on the investigation of textual representations of requirements engineering artifacts has been small. Eye tracking is mainly used to compare design alternatives in order to investigate visual effort and reading behavior. The use of eye tracking is ideal for investigating the three different linking variants between a use case and its associated requirements with respect to their visual effort and intended way of interrelating both artifacts.

## 4 Research Method

We aligned our study by following the recommendations for experimentation in software engineering by Wohlin et al. (2012). Thus, we applied the goal definition template to ensure that the important aspects of our experimental design are well-defined. Table 2 presents the goal definition for our eye tracking study.

**Table 2.** Goal definition

| | |
|---|---|
| *Object of study* | The three different linking variants between a use case and its associated requirements |
| *Purpose* | Evaluating their impact on the reading behavior |
| *Quality focus* | Visual effort and intended way of interrelating both artifacts |
| *Perspective* | Developers |
| *Context* | Using undergraduate and graduate students of computer science who need to process both artifacts and their interrelationships |

### 4.1 Experimental Design

Based on our goal definition, we selected a design science perspective resulting in a brain-based IT artifact evaluation. Figure 2 illustrates the abstract and concretized process of this evaluation approach by Riedl and Léger (2016).

This approach compares different design alternatives of an artifact to investigate their effect on a subject's behavior. The approach is based on the assumption that each design alternative causes a specific brain activity in terms of visual effort and cognitive load. This brain activity is the mediator which leads to a behavioral intention. This intention is the antecedent of the concrete observable behavior of a subject.

In our experiment, we focus on three different design alternatives for linking a use case with its associated requirements. We compare these alternatives each of which should cause a specific brain activity. The resulting brain activity should lead to the behavioral intention of switching between a use case and the requirements. We can observe and analyze the reading behavior of a subject by using eye tracking.

According to Sweller et al. (2011), cognitive load and visual effort depend not only on the design but also on the content. In this experiment, we are only interested in the impact of the different linking variants on reading behavior. For this reason, we decided to use the same use case and requirements for all subjects in order to exclude the effect of different contents. Therefore, we need a between-subjects design with 3 groups each of which uses only one of the three linking variants. A within-subjects design is not suitable since the subjects would know the contents after the first linking variant. Thus, they would have a learning effect for the other linking variants which would distort the results of the experiment.

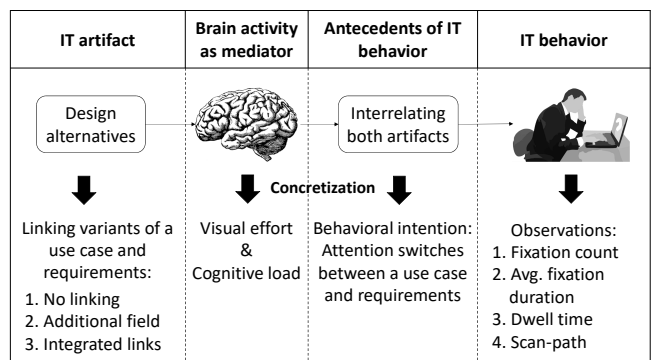

**Figure 2.** Brain-based IT artifact evaluation (Riedl and Léger, 2016)

### 4.2 Hypotheses

Based on our research question (see section 1), we test the following null hypotheses:

$H1_0$: There is no significant difference in the reading behavior of developers in terms of visual effort between the three linking variants while reading a use case and requirements to process both artifacts and their interrelationships.

$H2_0$: There is no significant difference in the reading behavior of developers in terms of the intended way of interrelating both artifacts between the three linking variants while reading a use case and requirements to process both artifacts and their interrelationships.

Each alternative hypothesis $Hi_1, i \in \{1, 2\}$ claims that the respective difference exists.

### 4.3 Material

We selected a specification of a software project that developed a software to maintain students, lecturers, and lectures of an institute of sinology at a different university. Specifically, we used the use case "Enter a lecturer" and a subset of 22 requirements (11 functional and 11 quality requirements). Five functional and 4 quality requirements were related to the use case. The use case follows the one-column template of Cockburn (2001, p. 121) with the 13 predefined fields. The authors of the use case have added an additional field for the links to the associated requirements. Based on this use case, we created the other two linking variants containing only the 13 predefined fields. Thus, all subjects got materials with the same content but different linking variants. Figure 3 shows exemplary snippets of the linking variants *additional field* (see Figure 3a) and *integrated links* (see Figure 3b) as well as some of the requirements used (see Figure 3c). The links are realized by the enumeration of the respective labels of the

**Figure 3.** Exemplary snippets of the used materials

(a) Snippet of the use case description with an *additional field*

| Technology | None |
|---|---|
| Associated requirements | [R101], [R103], [R107], [107.1], [R108], [R401], [R502], [R601], [R602] |

(b) Snippet of the use case description with *integrated links*

| Guarantee | The client system will not be affected. A consistent state of the database will be guaranteed. [R602] |
|---|---|
| Case of success | System saves the entered data of a lecturer in the database- |
| Trigger | User selects the function "Enter a lecturer" |
| Basic sequence | 1 | User selects the function "Enter a lecturer" |
| | 2 | System shows a window for entering the data |
| | 3 | User enters the data [R107], [R107.1], [R108] |

(c) Snippet of used requirements

| ID | Requirements |
|---|---|
| | **Functional requirements** |
| R101 | The system shall provide the user with the ability to enter a lecturer. |
| R102 | The system shall provide the user with the ability to delete a lecturer. |
| … | … |
| R107 | The data set of a lecturer shall at least consist of surname, forename, and gender. |
| R107.1 | The data set of one lecturer should contain additionally the degree, address, email address, phone number, consultation-hour, and room. |
| R108 | The data element gender has the characteristic f for female or m for male. |
| … | |
| | **Integrity** |
| R601 | The system shall require a login with user name and password to ensure that only authorized user access the data. |
| R602 | The database shall be consistent all time. |

associated requirements (see Figure 3, yellow markers). The variant *no linking* does not contain any links. We published all materials (guide sheet, requirements, use cases, and collected data) online to enable further replications and increase the transparency of our findings (Karras, 2021).

We created a PDF file consisting of two pages for each linking variant. The first page contains the particular use case and the second page contains the 22 requirements. We divided the use case and the requirements on separate pages due to the fact that our eye tracking system does not support scrolling and both artifacts on one page would have been unreadable. The eye tracking system gathers the data for each page separately. This technical constraint does not impact the determination of the visual effort. However, we needed to determine the scan-paths over both artifacts manually since our eye tracking system can only calculate a scan-path for a single page and not for a whole PDF document.

### 4.4 Independent and Dependent Variables

Our independent variable is the applied linking variant with three levels: *no linking*, *additional field*, and *integrated links*.

The dependent variables are the reading behavior in terms of visual effort and the intended way of interrelating both artifacts. We used three metrics for the visual effort: *fixation count*, the *average fixation duration*, and the *dwell time*. Besides the frequently used metrics *fixation count* and *average fixation duration* (Sharafi et al., 2015), we also decided to consider the *dwell time* since this time is the sum of the durations of fixations and saccades representing actual reading time due to eye movements.

For the intended way of interrelating both artifacts, we applied scan-path analysis. We focused on *attention switching frequency* between the use case and the requirements to describe the efforts to interrelate the two artifacts. The metric *attention switching frequency* is based on the idea that the more frequently a subject switches between both artifacts, the higher are his efforts to relate them (Sharafi et al., 2015). We also applied sequential pattern mining (Ayres et al., 2002) to identify *frequent sequential patterns* in the reading behavior. Table 3 summarizes all used metrics.

We gather the data about fixations, saccades, and scan-paths by using areas of interest (AOIs). An area of interest is a relevant element in the material. We define each field of a use case and each requirement as relevant and thus as an individual AOI to get detailed information about the subjects' reading behavior. In total, we have a set of 35 respectively 36 AOIs depending on the linking variant (see Table 4).

**Table 3.** Overview of the applied metrics

| Metrics for visual effort | |
|---|---|
| Fixation count | Number of all fixations in AOI |
| Avg. fixation duration | Avg. time of all fixations in AOI |
| Dwell time | Sum of durations from all fixations and saccades in AOI |
| **Metrics for intended way of interrelating** | |
| Attention switching frequency | Number of attention switches between AOIs |
| Frequent sequential patterns | Patterns of the frequent subsequences within all scan-paths |

**Table 4.** Number of defined AOIs per linking variant

| | **Number of AOIs** | | |
|---|---|---|---|
| **Linking variant** | Use case | Requirements | Total |
| *No linking* | 13 | 22 | 35 |
| *Additional field* | 14 | | 36 |
| *Integrated links* | 13 | | 35 |

The mitigating variable that might impact the effect of the independent variable on the dependent variables is the level of knowledge of our subjects. A subject's knowledge and experience are crucial concerns in an eye tracking study since they strongly influence the reading behavior and thus eye tracking data in terms of visual effort and cognitive load (Sweller et al., 2011). According to Sjøberg et al. (2002), the variations among professionals are generally higher than the variations among students due to a more varied educational background, working experience, etc. For this reason, we decided to limit the context of our goal definition to undergraduate and graduate students of computer science for the first experiments (see Table 2).

### 4.5 Subject Selection and Assignment

We used convenience sampling to select the subjects for our eye tracking study. All our subjects were undergraduate and graduate students of computer science at Leibniz University Hannover. Students are typical subjects in eye tracking studies. According to the results of a systematic literature review of eye tracking in software engineering (Sharafi et al., 2015), 25 out of 36 papers used students as subjects in their studies. First of all, the main reasons for using students as subjects are "*that they are more accessible and easier to organize, and hiring them is generally inexpensive*" (Sjøberg et al., 2002, p. 4). Furthermore, in contrast to professionals, students form a more homogeneous group since they have a comparable knowledge and are generally not very experienced. We checked these assumptions on knowledge and experience made for the experimental design by means of a questionnaire. A more homogeneous group reduces the effects on the eye tracking data by different factors such as knowledge, experience, and age (Sweller et al., 2011). As a consequence, the observed effects in the experiment can be clearly attributed to the linking variants. In several disciplines, self-selected students are considered as an appropriate subject pool for the study of social behavior (Exadaktylos et al., 2013), although experiments with students are often associated with a lack of realism (Sjøberg et al., 2002) and reduced external validity (Höst et al., 2000; Runeson, 2003).

All subjects participated voluntarily. There was no financial reward and thus little incentive to participate in our study without being self-motivated. We offered over 100 appointments on different days and times for the experiment sessions by using Doodle[2] to increase the willingness to participate. Thus, potential subjects could choose an individual appointment that fitted in their dairy. We used several communication channels to reach suitable subjects for the experiment. The Doodle was distributed via announcements in different computer science lectures, on our website, in an online forum of the computer science students of our university, and in the eLearning system of our university. As a consequence, the order of the subjects was completely random. This random order complicated the even distribution of the undergraduate and graduate students among the three groups. We tried to ensure that undergraduate and graduate students were distributed among the three groups as evenly as possible.

### 4.6 Experimental Setup

Figure 4 shows the setup for our experiment. We conducted all experiment sessions in a small, quiet room. The eye tracker (see Figure 4, 1) was placed below the 24-inch screen (see Figure 4, 2) with a resolution of $1920 \times 1280$ that was used to show the use case and the requirements to the subjects. Each subject was seated approximately 60cm away from the 24-inch screen in a chair with armrests but without wheels to maintain the same seating position during the whole session. The subject got a keyboard to navigate between the use case and the requirements. The experimenter sat on the left side of the subject with a 15-inch laptop (see Figure 4, 3) running the experimentation software. The experimenter used the laptop to conduct the experiment and to observe the subject. The 24-inch screen, the keyboard, and the eye tracker were connected to the laptop.

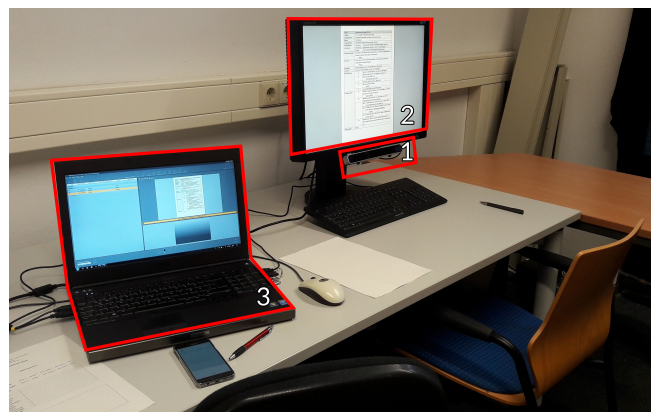

**Figure 4.** Setup – Eye tracker (1), 24-inch screen (2), and laptop (3)

### 4.7 Eye Tracking System

We used the RED250mobile eye tracker of SensoMotoric Instruments (SMI)[3] which supports fully automated image processing based on contact-free eye tracking and head movement compensation. It uses a binocular smart automatic tracking mode with a sampling rate of 250Hz to capture information such as fixations and saccades. The experiment planning, capturing, and analysis was done with the software provided by SMI. The *Experiment Center* software supports the design, planning, and execution of an eye tracking experiment. The *BeGaze* software provides functions to visualize and analyze the eye tracking data which can be exported in CSV files for further statistical analysis.

### 4.8 Experimental Procedure

We refined the experimental procedure iteratively through three rounds of pre-tests. In each pre-test, a computer science researcher, who has been working in the software engineering group at Leibniz Universität Hannover for more than three years, took part in the experimental procedure. After each pre-test, we discussed how the procedure could be improved. Below, we describe the final experimental procedure which consists of four steps: an introduction, a pre-questionnaire, the eye tracking, and a post-questionnaire.

Before running the experiment, each subject got a short introduction in which we briefly described the experimental procedure and the eye tracking system, e.g., how it works and which data is gathered. Afterward, the subject signed the declaration of consent to participate in our study.

With the pre-questionnaire, we gathered subjects' information such as passed computer science lectures and the level of knowledge about use cases. Based on the collected information, we were able to ensure that the subjects had the minimum knowledge required to be suitable for our experiment.

The eye tracking started with an explanation of the task of reading an excerpt from a specification. We provided a guide sheet with a task and scenario description. Each subject was put in the situation of being a developer in an ongoing project who has to process the excerpt of the specification to implement a use case. For each subject, we first calibrated the eye tracking system with a 5-point calibration. After the calibration, we started the experiment, presented the respective PDF file (see section 4.3), and collected the eye tracking data. During the eye tracking, the subject was free to decide how he reads and navigates between the use case and the requirements. We set a time limit of 10 minutes for the reading task. However, the subject could finish the reading task earlier if he believed that he had fulfilled the given task.

---

[2]https://www.doodle.com

[3]https://www.smivision.com

At the end of the experiment, the subject filled out a post-questionnaire. With the second questionnaire, we gathered qualitative information about a subject's reading behavior and decision. For example, we asked whether he had read specific requirements or not. This information should support the interpretation of the objective eye tracking data.

### 4.9 Data Analysis

The data analysis is composed of descriptive and inferential statistics. The data which is based on fixation count, average fixation duration, dwell time, and attention switching frequency (see Table 3) can be analyzed either with the parametric one-way *ANOVA* test or the non-parametric *Kruskal-Wallis* test. The choice of the test depends on whether the respective data is normally distributed or not. Data is normally distributed if the normality of residuals is met which can be tested by using the *Kolmogorov-Smirnov* and *Shapiro-Wilk* tests. In the case of normal distribution, we perform a one-way *ANOVA* test, and otherwise, a *Kruskal-Wallis* test, both with the significance level $p = 0.05$.

For the frequent sequential patterns, we apply sequential pattern mining using the *SPAM* algorithm (Ayres et al., 2002). This approach identifies frequent sequential patterns in the subjects' reading behavior. For this, we need the data in the form of scan-paths. A scan-path provides information on how a subject reads and switches between the use case and the requirements. In general, a scan-path is a sequence of visited AOIs arranged in chronological order. A frequent sequential pattern is a sub-sequence which appears in at least a specified minimum number of all analyzed scan-paths.

## 5 Baseline Experiment

### 5.1 Sample

The study subjects were 15 volunteers, 10 undergraduate and 5 graduate students of computer science. Table 5 summarizes the characteristics of the sample. The subjects were between the 3rd and 5th academic year, with an average academic year of 3.6. All of them were close to their graduation and had a similar basic level of knowledge with respect to use cases. We determined their level of knowledge based on the assessment of the statement "*I have a lot of experience with use cases.*" on a Likert scale ranging from "strongly disagree (1)" to "strongly agree (4)". Two subjects rated with "strongly disagree", 9 with "disagree", and 4 with "agree". The median level of knowledge was 2 for the undergraduates, 2 for the graduates, and 2 in total. Therefore, the subjects formed a homogeneous group. This consistent perspective helped us to encounter the mitigating variables since the level of knowledge has a strong influence on the reading behavior and thus the eye tracking data (Sweller et al., 2011).

**Table 5.** Baseline experiment: Key characteristics of subjects

| Characteristic | Level | # Subjects |
|---|---|---|
| Degree | Undergraduate | 10 |
| | Graduate | 5 |
| Academic year | 3rd | 7 |
| | 4th | 7 |
| | 5th | 1 |
| Level of knowledge: "I have a lot of experience with use cases." | Strongly disagree | 2 |
| | Disagree | 9 |
| | Agree | 4 |
| | Strongly agree | 0 |

### 5.2 Execution

We conducted each experiment session as scheduled according to the experimental procedure (see section 4.8). On average, all entire experiment sessions lasted 20 minutes. The minimum and maximum duration of all experiment sessions were 13 and 31 minutes, respectively. After having finished the data collection, we spent approximately 1.5 hours per experiment session to prepare, extract, and analyze the respective data record. The entire analysis required a total of approximately 22.5 hours.

### 5.3 Results

#### 5.3.1 Visual Effort

We used three metrics for visual effort calculation based on different eye tracking data: number of fixations, duration of fixations, and duration of fixations and saccades (see Table 3). Table 6 shows the calculated overall fixation count, overall average fixation duration [ms], and overall dwell time [min] of all AOIs of each subject in total. The results are grouped by the three linking variants.

For all three dependent variables, we checked whether the data is normally distributed by applying the *Kolmogorov-Smirnov* (KS) and *Shapiro-Wilk* (SW) test. In the case of a normal distribution, we performed a one-way *ANOVA* test otherwise a *Kruskal-Wallis* test on the respective data (see Table 6) with the significance level $p = 0.05$.

*1) Overall Fixation Count:* Both tests for normality of residuals showed that the overall fixation counts are not normally distributed (KS: $K = 0.24, p = 0.02$ and SW: $W = 0.83, p = 0.01$). We performed a *Kruskal-Wallis* test on the overall fixation counts (see Table 6). The test indicated that there is no statistically significant difference between the overall fixation counts by the linking variants ($\chi^2 = 0.74, p = 0.69, \eta^2 = 0.05$), with a mean rank of 8.8 for *no linking*, 6.6 for *additional field*, and 8.6 for *integrated links*. We cannot reject $H1_0$. Based on the overall fixation count, **there is no significant difference in the reading behavior in terms of visual effort between the three linking variants.**

*2) Overall Average Fixation Duration:* The *Kolmogorov-Smirnov* and *Shapiro-Wilk* tests indicated that the overall average fixation durations are normally distributed (KS: $K = 0.20, p = 0.12$ and SW: $W = 0.94, p = 0.32$). The overall average fixation duration (see Table 6) was analyzed with a one-way *ANOVA* test. The analysis showed that the effect of linking variant on the overall average fixation duration is not significant, $F(2, 12) = 1.51, p = 0.26, \eta^2 = 0.20$. $H1_0$ cannot be rejected. Based on the overall average fixation duration, **there is no significant difference in the reading behavior in terms of visual effort between the three linking variants.**

**Table 6.** Baseline experiment: Results for visual effort – fixation count, average fixation duration [ms], and dwell time [min]

| Treatment | Subject | Fixation count | Average fixation duration | Dwell time |
|---|---|---|---|---|
| No linking | P01 | 1200 | 137.9 | 03:21 |
| | P04 | 1095 | 131.1 | 03:07 |
| | P07 | 513 | 109.5 | 01:38 |
| | P10 | 943 | 147.7 | 02:46 |
| | P13 | 1129 | 141.9 | 03:16 |
| Additional field | P02 | 366 | 112.5 | 01:08 |
| | P05 | 711 | 164.5 | 02:14 |
| | P08 | 783 | 134.6 | 02:16 |
| | P11 | 1013 | 146.2 | 02:56 |
| | P14 | 2602 | 117.4 | 07:17 |
| Integrated links | P03 | 852 | 115.6 | 02:21 |
| | P06 | 1198 | 141.9 | 03:29 |
| | P09 | 893 | 104.1 | 02:47 |
| | P12 | 713 | 111.0 | 01:59 |
| | P15 | 1550 | 118.0 | 04:45 |

*3) Overall Dwell Time:* The tests for normality of residuals showed that the overall dwell times are not normally distributed (KS: $K = 0.24, p = 0.02$ and SW: $W = 0.83, p = 0.01$). We investigated the overall dwell time (Table 6) with a *Kruskal-Wallis* test. The test yielded no statistically significant difference between the overall dwell times by the linking variants ($\chi^2 = 0.56, p = 0.76, \eta^2 = 0.04$), with a mean rank of 8.4 for *no linking*, 6.8 for *additional field*, and 8.8 for *integrated links*. We cannot reject $H1_0$. Based on the overall dwell time, **there is no significant difference in the reading behavior in terms of visual effort between the three linking variants.**

**Finding $_{E-1}$**: The three linking variants do not differ with respect to the visual effort. Adding links to a use case does not impact the reading behavior in terms of visual effort.

#### 5.3.2 Intended Way of Interrelating Both Artifacts

Links between a use case and its associated requirements highlight their interrelationships. These links are intended to cause a specific reading behavior. A reader should follow the links to interrelate both related elements in order to process them. Reading behavior can be described based on a reader's scan-path. A scan-path is a sequence of visited AOIs arranged in the chronological order. Thus, scan-paths provide information on how the subjects read and switch between the use case and the requirements. We achieved very detailed scan-paths due to the 13 respectively 14 AOIs over the entire, respective use case and the 22 AOIs covering all requirements (see Table 4). We focused on attention switching frequency and sequential pattern mining to investigate the reading behavior.

*1) Attention Switching Frequency:* We used the attention switching frequency between all AOIs of the use case and requirements since the number of attention switches represents the subjects' efforts to interrelate both artifacts.

Figure 5 represents the attention switches between the use case and the requirements of each subject grouped by the linking variants. We also determined for each attention switch whether it interrelates the last considered use case field with its associated requirement (see Figure 5, black markers). For example, the first attention switch of subject P06 from the use case (dark gray) to the requirements (bright gray) was caused by a link (black marker). In the case of *no linking*, we considered whether the subjects switched from a use case field to the associated requirement on their own. The overall attention switching frequency of *integrated links* with a mean of 13.4 overall attention switches is larger compared to the other two linking variants. Whereas *no linking* has on average 2.0 overall attention switches, *additional field* has an average of 5.2 overall attention switches.

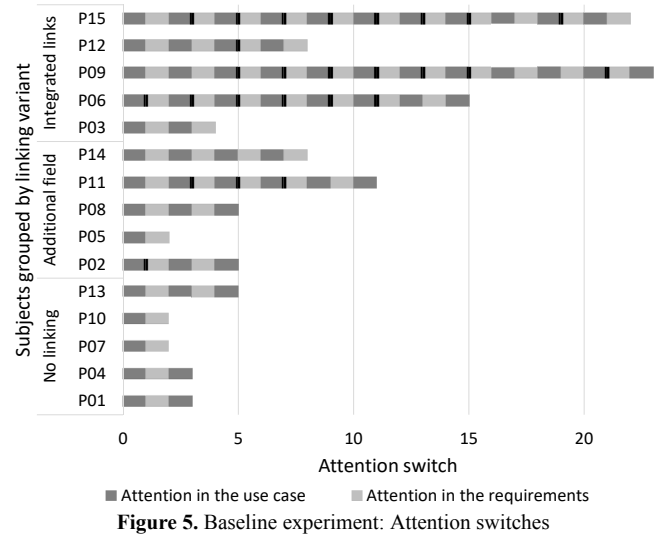

**Figure 5.** Baseline experiment: Attention switches

Considering the directed attention switching frequency from the use case to the requirements, the average number of attention switches of *integrated links* ($M = 7.0$) is larger than the average number of attention switches of the *additional field* variant ($M = 2.8$). The *no linking* group has the smallest average number of attention switches ($M = 1.2$).

We checked the directed attention switching frequency from the use case to the requirements for normal distribution with the *Kolmogorov-Smirnov* and *Shapiro-Wilk* test since the links were only defined in this direction. Both tests showed that the directed attention switching frequencies are not normally distributed (KS: $K = 0.29, p = 0.001$ and SW: $W = 0.76, p = 0.001$). Therefore, we performed a *Kruskal-Wallis* test with a significance level of $p = 0.05$. The test showed that there is a statistically significant difference between the directed attention switching frequencies by the linking variants ($\chi^2 = 8.21, p = 0.02, \eta^2 = 0.55$), with a mean rank of 3.9 for *no linking*, 8.1 for *additional field*, and 12.0 for *integrated links*. Hence, we can reject the null hypothesis $H2_0$. Based on the directed attention switching frequency, **there is a significant difference in the reading behavior in terms of the intended way of interrelating both artifacts.** According to Cohen (2013), the effect size ($\eta^2 = 0.55$) indicates a large practical relevance. The post-hoc pairwise comparison test using the *Bonferroni-Dunn* test showed that the mean score for the *integrated links* condition ($M = 7.0, SD = 4.1$) was significantly different from the *no linking* condition ($M = 1.2, SD = 0.4$). However, the *additional field* condition ($M = 2.8, SD = 1.6$) did not significantly differ from the *integrated links* and *no linking* conditions. Summarized, we can say that these results show that *integrated links* have a large effect on the reading behavior in terms of interrelating both artifacts. **Our results indicate that only the most detailed linking variant results in a reading behavior that includes interrelating both artifacts more intensively.**

We calculated the ratio between attention switches from a use case field to its associated requirements and all attention switches from the use case to the requirements. The subjects of the *no linking* variant achieved a ratio of $0.0\%$ since they did not match any use case field and associated requirement on their own. While the *additional field* variant resulted in an average ratio of $22.0\%$, the *integrated links* variant achieved an average ratio of $49.4\%$. In the case of *integrated links*, the links caused on average every second attention switch.

> **Finding $_{E-2}$**: Only the *integrated links* variant results in statistically significant more efforts by readers to interrelate the use case and the requirements. According to our results, these detailed links impact the reading behavior since they caused on average every second attention switch from the use case to the requirements.

*2) Frequent Sequential Patterns:* We applied sequential pattern mining (Ayres et al., 2002) on the scan-paths to identify the most frequent sequential patterns in the subjects' reading behavior.

Sequential pattern mining requires sequences of symbols from a fixed item set. However, if we use the 35 respectively 36 AOIs (see Table 4) as an item set in the case of only 5 subjects per group the resulting sequences of the captured scan-paths are too divergent. As a consequence, they do not share any sequential pattern with a sufficient support by the sequences. Therefore, we decided to simplify the sequences by defining a smaller item set. We defined the item set $I = \{UCR, RQR, INT\}$ that describes the possible reading options. Either the subject reads in the use case (UCR) respectively in the requirements (RQR) or the subject interrelates (INT) both artifacts to process their interrelationships. Three raters classified each visited AOI of each scan-path on their own as one of the three options. This classification could not be done by our subjects since they were no longer available at the time of the data analysis. We evaluated the reliability of the raters' classification by using *Fleiss' kappa* (Fleiss, 1971). *Fleiss' kappa* is a measure of the agreement between a fixed number of raters greater than two, where agreement due to chance is factored out. The calculated *Fleiss' kappa* value was 0.85 which shows an almost perfect raters' agreement according to Landis and Koch (1977). We used three raters to achieve for each AOI a majority decision since an AOI can only be either UCR respectively RQR or INT. Based on the classification, we derived simplified sequences by aggregating successive AOIs with the same label as one symbol of the respective label similar to Uwano et al. (2006).

Figure 6 shows the resulting sequences which we used for the sequential pattern mining. For example, subject P12 reads the use case, then the requirements, and interrelates both artifacts at the end. The resulting sequence is encoded as $\langle UCR, RQR, INT \rangle$.

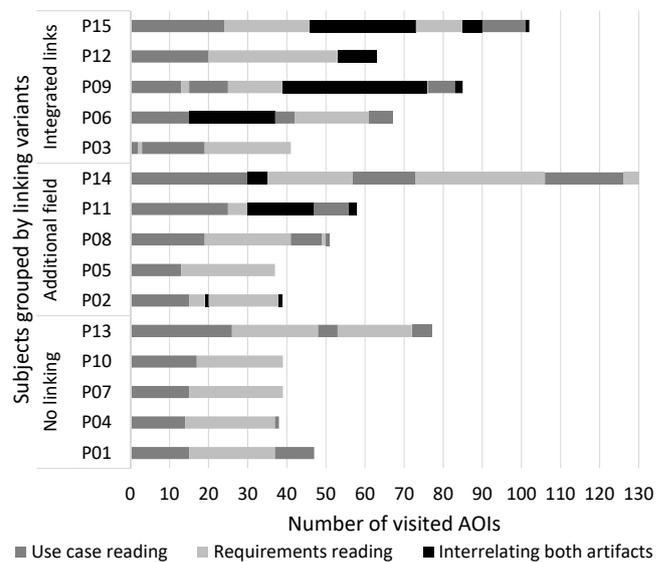

**Figure 6.** Baseline experiment: Simplified sequences of scan-paths

We performed the sequential pattern mining on the sequences presented in Figure 6 by using the *SPAM* algorithm (Ayres et al., 2002) for each linking variant. A sequential pattern is a sub-sequence which appears in at least a specified minimum number of sequences. These sequences support the identified sequential pattern. We selected a minimal support of 3 which means we decided that a frequent sequential pattern should appear in more than $50\%$ of the analyzed sequences. Figure 7 shows the identified frequent sequential reading patterns grouped by the linking variants. All three linking variants share the most frequent sequential reading pattern $\langle UCR, RQR \rangle$ with a support of 5 per group and an overall frequency of 20. Interrelating both artifacts (INT) only appears as part of sequential reading patterns of the *integrated links* group in the sub-sequences $\langle UCR, RQR, INT \rangle$, $\langle INT, UCR \rangle$, and $\langle RQR, INT \rangle$ all with a support of 3 and a frequency of 3 respectively 4. The most frequent sequential reading pattern that is shared by all linking variants is reading the use case and the requirements separately and successively. Interrelating both artifacts only occurred in the case of visualized links. However, we have to restrict this finding by emphasizing that interrelating the two artifacts is only part of frequent sequential patterns of *integrated links*.

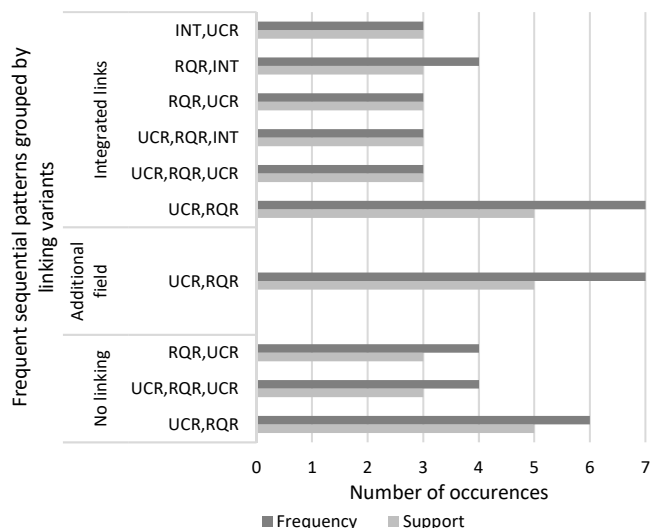

**Figure 7.** Baseline experiment: Frequent sequential reading patterns

**Finding $_{E-3}$**: The main reading behavior of all three linking variants is the successive reading of the single artifacts. The intended way of interrelating both artifacts only occurred frequently in the case of *integrated links*.

# 6 Repetition

Below, we describe the replication of the experiment. In particular, this replication is a repetition to control potential sampling errors. This procedure is the first necessary step to replicate experiments according to the systematic approach by Gómez et al. (2014). In this way, we can verify that the results of the baseline experiment (Karras et al., 2018) are not a product of chance. For this purpose, we repeated the exact same experiment with 15 new subjects of the same population. We were able to balance the uneven distribution of undergraduate and graduate students among the three groups. We spent more efforts on subject selection and assignment to achieve complementary groups compared to the baseline experiment. As a side effect, we increased the sample size through repetition by having the same total number of 5 undergraduate and 5 graduate students per group. This larger sample of 30 subjects can be analyzed in total since the two experiments follow the exact same research method (see section 4). Therefore, the joint analysis of both experiments contributes to an increased conclusion validity of the findings.

## 6.1 Sample

The subjects were 15 new volunteers, 5 undergraduate and 10 graduate students of computer science. Table 7 presents the key characteristics of the sample. The subjects were between the 3rd and 6th academic year, with an average academic year of 4.2. All of them had a similar basic level of knowledge with respect to use case like the sample of the baseline experiment. Regarding the statement "*I have a lot of experience with use cases.*", Two subjects rated with "strongly disagree", 11 with "disagree", and 2 with "agree". The median level of knowledge was 2 for the undergraduates, for the graduates, and in total. Thus, the subjects formed a homogeneous group.

**Table 7.** Repetition: Key characteristics of subjects

| Characteristic | Level | # Subjects |
|---|---|---|
| Degree | Undergraduate | 5 |
| | Graduate | 10 |
| Academic year | 3rd | 4 |
| | 4th | 6 |
| | 5th | 3 |
| | 6th | 2 |
| Level of knowledge: "I have a lot of experience with use cases." | Strongly disagree | 2 |
| | Disagree | 11 |
| | Agree | 2 |
| | Strongly agree | 0 |

## 6.2 Execution

We conducted the repetition according to the experimental procedure (see section 4.8). While the minimum and maximum durations of all experimental sessions were 10 respectively 26 minutes, the entire experimental procedure (see section 4.8) lasted on average 19 minutes. The data analysis required a total of approximately 22.5 hours since we needed approximately 1.5 hours for the preparation, extraction, analysis of a single data record of one experiment session.

## 6.3 Results

### 6.3.1 Visual Effort

Table 8 presents the overall fixation count, overall average fixation duration [ms], and overall dwell time [min] of all AOIs in the use case and the requirements of each subject.

We checked whether the data of all three dependent variables is normally distributed by using the *Kolmogorov-Smirnov* (KS) and *Shapiro-Wilk* (SW) test. In the case of normal distribution, we performed a one-way *ANOVA* test, otherwise a *Kruskal-Wallis* test on the respective data with a significance level of $p = 0.05$.

*1) Overall Fixation Count:* While the *Kolmogorov-Smirnov* test indicated that the overall fixation counts are normally distributed (KS: $K = 0.20, p = 0.09$), the *Shapiro-Wilk* test showed that the data is not normally distributed (SW: $W = 0.82, p = 0.007$). We decided to perform the non-parametric *Kruskal-Wallis* test due to the uncertainty regarding the normal distribution of the data. The test yielded that there is no statistically significant difference between the overall fixation counts by the three linking variants ($\chi^2 = 0.11, p = 0.95, \eta^2 = 0.01$), with a mean rank of 8.4 for *no linking*, 7.5 for *additional field*, and 8.1 for *integrated links*. Hence, we cannot reject the null hypothesis $H1_0$. Based on the overall fixation count, **there is no significant difference in the reading behavior in terms of visual effort between the three linking variants.**

*2) Overall Average Fixation Duration:* The two tests for normality of residuals showed that the overall average fixation durations are normally distributed (KS: $K = 0.18, p = 0.22$ and SW: $W = 0.94, p = 0.34$). We performed a one-way *ANOVA* test. The analysis indicated that the effect of linking variant on the overall fixation duration is not significant,

Table 8. Repetition: Results for visual effort – fixation count, average fixation duration [ms], and dwell time [min]

| Treatment | Subject | Fixation count | Average fixation duration | Dwell time |
|---|---|---|---|---|
| No linking | P16 | 943 | 157.7 | 02:54 |
| | P19 | 1100 | 138.2 | 03:12 |
| | P22 | 1276 | 122.0 | 03:21 |
| | P25 | 1241 | 112.7 | 03:28 |
| | P28 | 868 | 128.7 | 02:17 |
| Additional field | P17 | 686 | 126.8 | 01:56 |
| | P20 | 679 | 119.5 | 01:45 |
| | P23 | 1299 | 136.8 | 03:33 |
| | P26 | 1319 | 123.1 | 03:21 |
| | P29 | 965 | 122.6 | 02:29 |
| Integrated links | P18 | 702 | 100.3 | 02:12 |
| | P21 | 965 | 155.2 | 02:55 |
| | P24 | 2213 | 169.9 | 07:08 |
| | P27 | 925 | 102.3 | 02:36 |
| | P30 | 1128 | 123.5 | 03:07 |

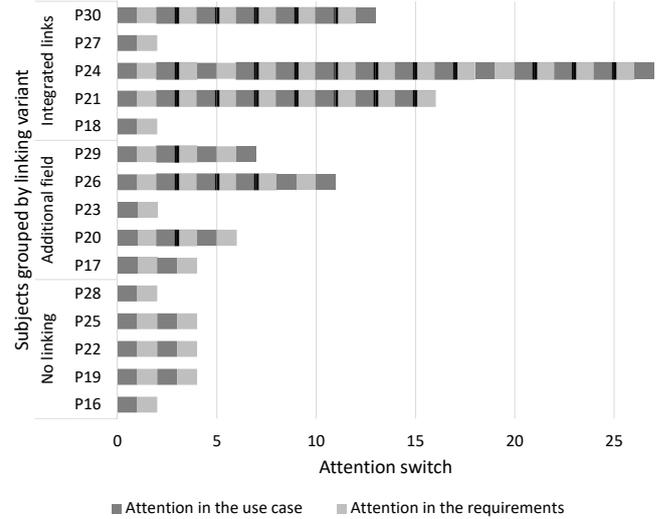

Figure 8. Repetition: Attention switches

$F(2, 12) = 0.11, p = 0.89, \eta^2 = 0.02$. We cannot reject $H1_0$. Based on the overall fixation duration, **there is no significant difference in the reading behavior in terms of visual effort between the three linking variants.**

*3) Overall Dwell Time:* The *Kolmogorov-Smirnov* and *Shapiro-Wilk* test indicated that the overall dwell times are not normally distributed (KS: $K = 0.29, p = 0.002$ and SW: $W = 0.72, p < 0.001$). We performed a *Kruskal-Wallis* test. The test showed no statistically significant difference between the overall dwell times by the linking variants ($\chi^2 = 0.86, p = 0.65, \eta^2 = 0.06$), with a mean rank of 9.2 for *no linking*, 6.6 for *additional field*, and 8.2 for *integrated links*. $H1_0$ cannot be rejected. Based on the overall dwell time, **there is no significant difference in the reading behavior in terms of visual effort between the three linking variants.**

**Finding** $_{R-1}$: As in the baseline experiment, adding links to a use case does not have an impact on the reading behavior in terms of visual effort.

### 6.3.2 Intended Way of Interrelating Both Artifacts

*1) Attention Switching Frequency:* Figure 8 presents the attention switches between the use case and the requirements of each subject. We determined for each attention switch whether it interrelates the last considered use case field with its associated requirement (see Figure 8, black markers). In the case of the variant *no linking*, we analyzed whether the subjects switched from a use case field to its associated requirement on their own. The subjects of the *integrated links* variant switched overall on average 12.2 times. This mean is larger compared to the means of the other two linking variants. While the subjects of the *no linking* variant switched on average 2.1 times, the subjects of the *additional field* variant switched on average 5.1 times.

When we consider the directed attention switching frequency from the use case to the requirements, the average number of attention switches of the *integrated links* ($M = 6.4$) is larger than the average number of attention switches of the *additional field* variant ($M = 2.8$). The group of the *no linking* variant has the smallest average number of attention switches ($M = 1.2$).

We applied the *Kolmogorov-Smirnov* and *Shapiro-Wilk* test on the directed attention switching frequencies to check them for normality of residuals. Both tests showed that the data is not normally distributed (KS: $K = 0.28, p = 0.002$ and SW: $W = 0.74, p = 0.001$). We performed a *Kruskal-Wallis* test with a significance level $p = 0.05$. According to the results, there is no significant difference between the directed attention switching frequencies by the linking variants ($\chi^2 = 2.09, p = 0.35, \eta^2 = 0.34$), with a mean rank of 5.7 for *no linking*, 8.7 for *additional field*, and 9.6 for *integrated links*. $H2_0$ cannot be rejected. Based on the directed attention switching frequency, **there is no significant difference in the reading behavior in terms of the intended way of interrelating both artifacts.**

We calculated the ratio between attention switches from a use case field to its associated requirements and from the use case to the requirements. In the case of *no linking*, the subjects achieved a ratio of $0.0\%$ since they did not match any use case field and associated requirement on their own. The *additional field* variant achieved an average ratio of $25.3\%$ and the *integrated links* variant resulted in an average ratio of $49.6\%$. Thus, the *integrated links* caused on average every second switch.

**Finding** $_{R-2}$: In contrast to the baseline experiment, the three linking variants do not differ in the readers' efforts to interrelate both artifacts. Nevertheless, the *integrated links* variant also caused on average every second attention switch from the use case to the requirements.

*2) Frequent Sequential Patterns:* As with the baseline experiment, we needed to simplify the detailed scan-paths which were based on the 35 respectively 36 AOIs (see Table 4) in order to apply sequential pattern mining. We asked the same three raters to classify each visited AOI of each scan-path on their own as one of the three options: Reading in the use case (UCR), reading in the requirements (RQR), or interrelating both artifacts (INT) (see section 5.3.2, 2) *Frequent Sequential Patterns*). We evaluated the reliability of the raters' classification by using *Fleiss' kappa* (Fleiss, 1971). The calculated *Fleiss' kappa* value was 0.89 which represents an almost perfect agreement of the raters according to Landis and Koch (1977). We achieved for each AOI a majority decision due to the three raters since an AOI can only be either UCR respectively RQR or INT. The simplified sequences resulted from aggregating successive AOIs with the same label as one symbol the respective label. In Figure 9, we present the resulting simplified sequences which we used for the sequential pattern mining.

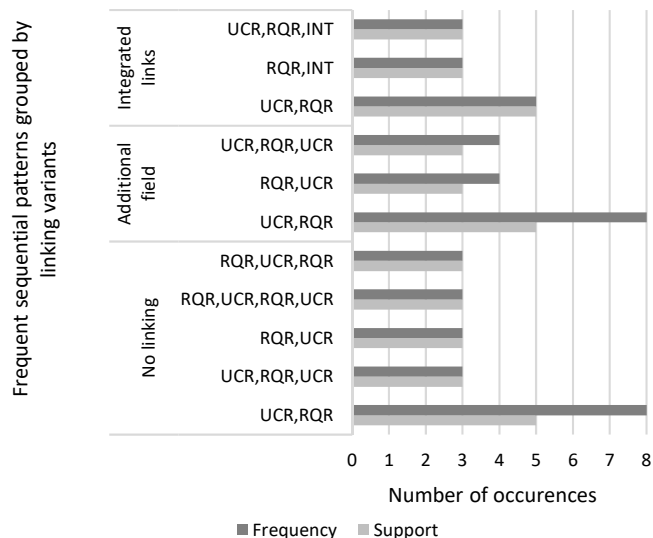

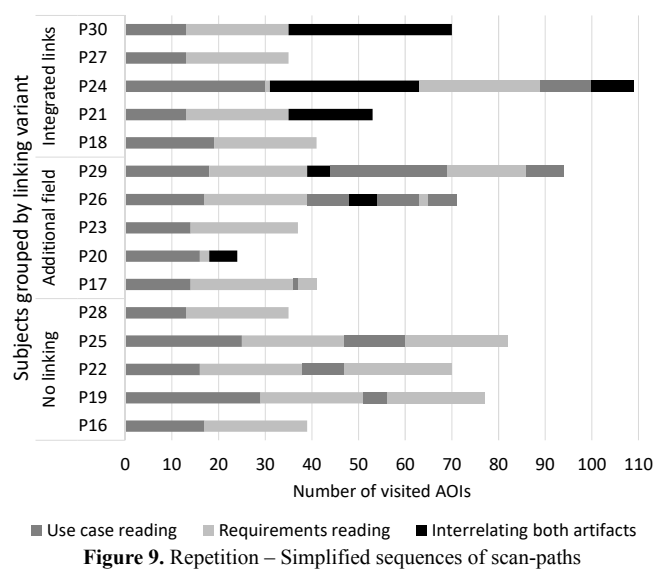

**Figure 9.** Repetition – Simplified sequences of scan-paths

We applied the SPAM algorithm (Ayres et al., 2002) on the simplified sequences for each linking variant. We selected 3 as a minimal support of a pattern by the sequences which means we decided that a frequent sequential pattern should appear in more than 50% of the analyzed sequences. Figure 10 summarizes all identified frequent sequential reading patterns grouped by the linking variants. The most frequent sequential reading pattern $\langle UCR, RQR \rangle$ is shared by all linking variants with a support of 5 per group and an overall frequency of 21. The reading behavior of interrelating both artifacts (INT) only occurs as parts of the *integrated links* variant in the subsequences $\langle UCR, RQR, INT \rangle$ and $\langle RQR, INT \rangle$ both with a support of 3 and a frequency of 3. As the most frequent sequential pattern, all three linking variants shared reading the use case and the requirements separately and successively. The reading behavior of interrelating both artifacts occurred only in the case of visualized links. This finding is restricted since only the frequent sequential patterns of *integrated links* contain this reading behavior.

**Finding $_{R-3}$**: As in the baseline experiment, all three linking variants share as the most frequent reading behavior the successive reading of the single artifacts. Only the *integrated links* variant contained in its frequent patterns the intended way of interrelating both artifacts.

**Figure 10.** Repetition: Frequent sequential reading patterns

## 7 Comparison of both Experiments

In this section, we compare the findings of the baseline experiment and its repetition. Rather than extracting only conclusions from the single experiments, we also conduct a joint analysis of both since they can be considered on the whole as one single experiment. Thus, we are able to: (a) Cancel out potential effects of the uneven distribution of undergraduate and graduate students by having overall 5 undergraduate and 5 graduate students per group; (b) Increase the conclusion validity of our findings. Table 9 summarizes the findings of the baseline experiment, the repetition, and the joint analysis.

### 7.1 Consistent Findings

Almost all findings of the baseline experiment and the corresponding findings of the repetition are consistent. The statistical analysis of the visual effort in terms of fixation count, average fixation duration, and dwell time shows completely consistent results. All statistical tests indicated no significant difference in the reading behavior in terms of visual effort between the three linking variants. Thus, the respective linking variant has no impact on the visual effort.

The sequential pattern mining also leads to the same findings for the baseline experiment as for the repetition. While reading the use case first and then the requirements $\langle UCR, RQR \rangle$ is the most frequent sequential pattern of all three linking variants, only the frequent sequential patterns of the *integrated links* variant contained the reading behavior of interrelating both artifacts (INT). Therefore, the typical reading behavior of all variants is the successive reading of the single artifacts. Only the *integrated links* variant leads to the intended way of interrelating both artifacts.

### 7.2 Inconsistent Findings

The only inconsistent findings between the baseline experiment and the repetition occurred in the statistical analysis of the attention switching frequency. The number of attention switches represents the subjects' efforts to interrelate both artifacts. While we identified a statistically significant difference in the number of directed attention switches between the *no linking* and *integrated links* variant in the baseline experiment, we did not find any difference in the repetition.

**Table 9.** Comparison of the baseline experiment, the repetition, and the joint analysis (Remark: *No Linking* := NL, *Integrated links* := IL)

| | **Metric** | **Baseline experiment** | **Repetition** | **Joint analysis** |
|---|---|---|---|---|
| Statistical analysis | Significant difference between the three linking variants with respect to | | | |
| | Fixation count | No | No | No |
| | Average fixation duration | No | No | No |
| | Dwell time | No | No | No |
| | Attention switching frequency | Yes, between NL & IL | No | Yes, between NL & IL |
| Sequential pattern mining | All three linking variants share the following most frequent sequential pattern | | | |
| | Frequent sequential patterns | $\langle UCR, RQR \rangle$ | $\langle UCR, RQR \rangle$ | $\langle UCR, RQR \rangle$ |
| | Interrelating both artifacts (INT) occurred in the frequent sequential patterns of | | | |
| | Frequent sequential patterns | Only IL | Only IL | Only IL |

The only difference between the two experiments is the composition of the groups. In the baseline experiment, we had difficulties to evenly balance the three groups since we had 10 undergraduate and 5 graduate students. As a consequence, we decided to form exact complementary groups compared to the baseline experiment by selecting 5 undergraduate and 10 graduate students as subjects in the repetition. Although all subjects had a similar level of knowledge about use cases (see Table 5 and Table 7), the progress in the study might have an effect on the reading behavior. We canceled out this potential effect in the joint analysis due to the complementary groups in the repetition since all three groups contained the same number of 5 undergraduate and 5 graduate students.

## 7.3 Joint Analysis

We analyzed the collected data of the two individual experiments together since we completely replicated the exact same research method with 15 new subjects. Therefore, the individual experiments can be considered as one experiment with a larger sample size. For a better comparison, we report the results of the joint analysis of all 30 subjects in the same structure as the results of the single experiments but more briefly.

### 7.3.1 Visual Effort

*1) Overall Fixation Count:* We checked the data (see Table 6 and Table 8) for normal distribution with the *Kolmogorov-Smirnov* (KS) and *Shapiro Wilk* (SW) test. Both tests showed that the overall fixation counts are not normally distributed (KS: $K = 0.18, p = 0.01$ and SW: $W = 0.84, p < 0.001$). Thus, we performed a *Kruskal-Wallis* test which indicated that there is no statistically significant difference between the overall fixation counts by the different linking variants ($\chi^2 = 0.66, p = 0.72, \eta^2 = 0.02$) with a mean rank of 16.9 for *no linking*, 13.8 for *additional field*, and 8.6 for *integrated links*. We cannot reject $H1_0$. Based on the overall fixation count, **there is no significant difference between in the reading behavior in terms of visual effort between the three linking variants.**

*2) Overall Average Fixation Duration:* According to both tests for normality of residuals, the overall average fixation durations (see Table 6 and Table 8) are normally distributed (KS: $K = 0.12, p = 0.32$ and SW: $W = 0.96, p = 0.40$). The applied one-way *ANOVA* test showed that the effect of linking variant on the overall average fixation duration is not significant $F(2, 27) = 0.56, p = 0.58, \eta^2 = 0.04$. $H1_0$ cannot be rejected. Based on the overall average fixation duration, **there is no significant difference between the reading behavior in terms of visual effort between the three linking variants.**

*3) Overall Dwell Time:* The tests for normality of residuals showed that the overall dwell times (see Table 6 and Table 8) are not normally distributed (KS: $K = 0.26, p < 0.001$ and SW: $W = 0.78, p < 0.001$). We performed a *Kruskal-Wallis* test that yielded no statistically significant difference between the overall dwell times by the different linkings variants ($\chi^2 = 1.23, p = 0.54, \eta^2 = 0.04$) with a mean rank of 17 for *no linking*, 13 for *additional field*, and 16.5 for *integrated links*. $H1_0$ cannot be rejected. Based on the overall dwell time, **there is no significant difference in the reading behavior in terms of visual effort between the three linking variants.**

> **Finding $_{JA-1}$**: The joint analysis confirms the results of both experiments. All linking variants do not differ regarding visual effort and thus do not have an impact on reading behavior in terms of visual effort.

### 7.3.2 Intended Way of Interrelating Both Artifacts

*1) Attention Switching Frequency:* According to the *Kolmogorov-Smirnov* and *Shapiro Wilk* tests, the directed attention switching frequencies (see Figure 5 and Figure 8) are not normally distributed (KS: $K = 0.28, p < 0.001$ and SW: $W = 0.76, p < 0.001$). We applied a *Kruskal-Wallis* test. The test showed that there is a statistically significant difference between the directed attention switching frequencies by the different linking variants ($\chi^2 = 10.11, p = 0.006, \eta^2 = 0.34$), with a mean rank of 9.1 for *no linking*, 16.2 for *additional field*, and 21.2 for *integrated links*. $H2_0$ can be rejected. Based on the directed attention switching frequency, **there is a significant difference in the reading behavior in terms of the intended way of interrelating both artifacts.** The effect size ($\eta^2 = 0.34$) indicates a large practical relevance (Cohen, 2013). A post-hoc pairwise comparison test using *Bonferroni-Dunn* test yielded that the mean score for the *integrated links* condition ($M = 6.4, SD = 4.4$) was significantly different from the *no linking* condition ($M = 1.4, SD = 0.5$). The *additional field* condition ($M = 2.8, SD = 1.5$) did not differ from any of the two other conditions. **The *integrated links* variant has a large effect on the reading behavior resulting in interrelating both artifacts more intensively.**

The ratio between attention switches from a use case to its associated requirements and all attention switches from the use case to the requirements was $0.0\%$ for the *no linking* variant, $23.7\%$ for the *additional field* variant, and $49.5\%$ for the *integrated links* variant. While the subjects of the *no linking* variant did not match any use case field and associated requirements on their own, the subjects of *integrated links* switched on average every second time due to the links.

> **Finding $_{JA-2}$**: The joint analysis clarifies the inconsistent findings between both experiments by confirming the findings of the baseline experiment. The *integrated links* variant impacts the reading behavior by leading to statistically significant more efforts to interrelate both artifacts.

*2) Frequent Sequential Patterns:* We performed the SPAM algorithm (Ayres et al., 2002) on the simplified sequences (see Figure 6 and Figure 9) for each linking variant with the same settings as in the two single experiments. Since we decided for the single experiments that a frequent sequential pattern should appear in more than $50\%$ of the analyzed sequences, the minimal support for the SPAM algorithm of the joint analysis is 6. In Figure 11, we present the identified frequent sequential patterns. $\langle UCR, RQR \rangle$ is again the most frequent sequential pattern shared by all three linking variants with a support of 10 per group and an overall frequency of 41. Interrelating both artifacts (INT) is only part of frequent sequential patterns of the *integrated links* variant in the subsequences $\langle UCR, RQR, INT \rangle$ and $\langle RQR, INT \rangle$ both with a support of 6 and a frequency of 6 respectively 7. Reading the single artifacts one after the other is the most typical behavior. The necessary behavior of interrelating both artifacts to process them only occurs in the case of *integrated links*.

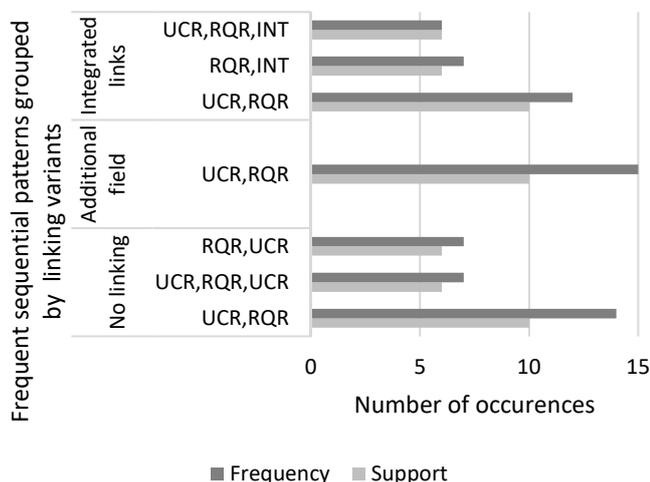

**Figure 11.** Joint Analysis – Frequent sequential reading patterns

> **Finding $_{JA-3}$**: The joint analysis confirms the results of both experiments. While all three linking variants share the successive reading of the single artifacts as the most frequent reading behavior, the intended way of interrelating both artifacts only occurred frequently in the case of *integrated links*.

## 7.4 Interpretation

The findings of the baseline experiment, the repetition, and the joint analysis are almost completely consistent. The only inconsistent finding regarding attention switching frequency was clarified by the joint analysis which confirmed the baseline experiment (see Table 9).

Our findings provide interesting insights with respect to the impact of the three linking variants on reading behavior. All linking variants cause a comparable visual effort and share the most frequent sequential pattern of reading the use case and the requirements one after the other. However, only the *integrated links* result in the intended way of interrelating both artifacts according to the directed attention switching frequency and frequent sequential patterns.

The reading behavior in terms of visual effort does not differ between the three linking variants. All three analyzed metrics for visual effort, which are based on different measures, show no significant difference. Thus, adding links consisting of unique and persistent identifiers that correspond to labels of requirements does not increase a reader's visual effort. This finding is plausible since all three linking variants result in visual representations of a use case that differ only slightly. These minimal visual differences caused by the links do not impede the perception and thus the reading of the artifacts.

Although there is no difference in the visual effort, we identified a significant difference in the intention to interrelate both artifacts. The overall attention switching frequency increases in the case of visualized links compared to *no linking*. This finding, however, is restricted since we only identified a statistically significant difference of the directed attention switching frequency between the *integrated links* and *no linking* condition. Based on the joint analysis, the *integrated links* variant ($M = 6.4$) leads to $78.1\%$ more directed attention switches from the use case to the requirements than the *no linking* variant ($M = 1.4$). **Therefore, the *integrated links* variant is the only linking variant that results in more efforts to interrelate both artifacts by a reader.** The sequential pattern analyses support this finding since we only identified interrelating both artifacts as a part of the frequent sequential patterns of the *integrated links* variant in the baseline experiment, repetition, and joint analysis. Hence, the intended way of interrelating both artifacts is only verifiable in the case of the most detailed linking. As an answer to our research question, we can summarize:

> **Answer to research question:**
> The three linking variants have no impact on reading behavior in terms of visual effort. They also share the most frequent sequential pattern of reading the single artifacts successively. Only the *integrated links* variant results in a reading behavior that includes interrelating both artifacts.

# 8 Threats to Validity

We considered threats to construct, external, internal, and conclusion validity according to Wohlin et al. (2012).

## 8.1 Construct Validity

Threats to construct validity address influences on the generalizability of an experiment's results to the concept or theory behind an experiment (Wohlin et al., 2012).

We used the same use case and requirements for all three linking variants. Thus, we had a mono-operation bias since we only used one dataset for the material of our experiment. As a consequence, the used material did not convey a comprehensive overview of the complexity in practice. Nevertheless, the material was from a real software project. Therefore, we expected that the selected material represented a sufficient realistic complexity for the subjects. We accepted this threat of a mono-operation bias to achieve a better comparability of our measurements. The analysis of reading behavior required an exact measuring. However, people are afraid of being evaluated and thus they are inclined to convey the impression of being better than they really are. This human tendency endangered the outcome of our experiment. We counteracted this threat to validity by using eye tracking for objective measurements of the subjects' behavior beyond doubt. Particularly, we used a contact-free eye tracker which can compensate for head movements to mitigate the presence of the eye tracker. As a result, we created a more natural behavior by our subjects since they sat as usual in front of a monitor while working with digital documents. However, the eye tracking system used affected the results since the use case and the requirements had to be divided into separate pages in a PDF file. We assume that we found $\langle UCR, RQR \rangle$ (reading the use case first and then the requirements) as the most frequent sequential pattern since the use case was presented on the first page and the requirements on the second page of the respective PDF file. As a consequence, we expect that we would have found $\langle RQR, UCR \rangle$ as the most frequent sequential pattern if we had swapped the order of the use case and the requirements. Nevertheless, this assumption does not change the overall finding that the most frequent sequential pattern is reading single artifacts successively. The single use of eye tracking caused a mono-method bias. All measurements were based on the eye tracking data and thus only allowed a restricted explanation of our findings. We mitigated this threat by using multiple measures for visual effort and interrelating both artifacts in order to cross-check them. We focused on these objective measures instead of subjective ones since objectives measures are easier to reproduce and thus more reliable. We also used a post-questionnaire for subjective measures. However, these answers were not as convincing as the eye tracking data since the answers partially contradicted themselves and the eye tracking data. The subjects did not remember the rationales behind their behavior and decisions due to the time between reading the artifacts and answering the post-questionnaire. We were unable to use the recorded eye tracking data immediately after the eye tracking session to help the subject to remember his behavior and decision since we would have needed approximately 45 minutes to prepare the material for consideration. Such a long waiting time could have impacted the subjects' motivation negatively, and since the time of the subjects is valuable, we decided not to waste it. We were also unable to resolve the contradictions retrospectively since the subjects were no longer available at the time of the analysis. The given task of reading an excerpt of a specification in combination with an eye tracker caused an interaction of testing and treatment. The use of an eye tracker and reading materials implied to analyze the reading behavior. The subject could have been aware of how to read the artifacts. This threat to validity was difficult to mitigate. However, we did not mention the existence and comparison of different linking variants. Thus, the subjects' attention was not focused on the links and each subject decided to use the links on his own.

## 8.2 External Validity

Threats to external validity address issues that limit the ability to generalize the results of an experiment to industrial practice (Wohlin et al., 2012).

The selected subjects, as well as the data from a real software project, produced a good level of realism. The undergraduate and graduate students were close to their graduation. Thus, they represented future developers which are one major role that intensively works with use cases and requirements. Our subjects formed a homogeneous group due to a similar level of knowledge, experience, and age. However, the subjects' homogeneity restricted the generalizability of our results. The subjects' level of knowledge about use cases might not be representative of developers from industry. We decided to accept this threat to validity in order to reduce potential effects on the eye tracking data due to differences in knowledge, experience, and age. These factors have a strong impact on the cognitive load, visual effort, and reading behavior. By canceling out these factors, the observed effects in the experiment could be clearly attributed to the linking variants. The artificial environment and the experimental setting also endangered the external validity. Processing the artifacts had no true pragmatic value for the subjects since none of them had a genuine working task. Future replications of this eye tracking study and similar evaluations should be done on real industry projects with developers and other roles that work with these artifacts.

## 8.3 Internal Validity

Threats to internal validity address influences that have a causal effect on the independent variable without the researcher's knowledge. These threats affect the conclusions about possible causal relationships between the treatments and the outcome of an experiment (Wohlin et al., 2012).

We had three different groups due to the selected between-subjects design. These groups caused interactions with selection since different groups have a different behavior. We consciously decided to apply this design in order to use the same material for all groups. Thus, we counteracted possible learning effects. Furthermore, we had difficulties to evenly distribute the undergraduate and graduate students in the baseline experiment. We formed complementary groups compared to the baseline experiment in the repetition. As a consequence, we achieved balanced groups consisting of 5 undergraduate and 5 graduate students for the joint analysis to cancel out potential effects. The selected between-subjects design also reduced the effort for our subjects since eye tracking is time-consuming and exhausting. A single session with one subject required as much as 20 minutes for one particular linking variant. A session with all three linking variants was not reasonable due to the following three reasons. First, we would need different but comparable use cases and requirements to avoid learning effects. Second, the subjects would need to keep a steady sitting position for over one hour for comparable eye tracking results. Third, the resulting eye tracking data of one subject would require approximately 4.5 hours for preparation, extraction, and analysis resulting in 135 hours for the whole data analysis of 30 subjects.

## 8.4 Conclusion Validity

These threats to validity address issues that affect the ability to draw conclusions about relations between the treatments and the outcome of an experiment (Wohlin et al., 2012).

We used eye tracking to improve the reliability of our results since the use of objective measures is easier to reproduce and more reliable than subjective ones. However, eye tracking data, visual effort, and cognitive load are influenced by different factors such as knowledge, experience, and age which are difficult to control. We counteracted this threat to validity by only selecting students as subjects that were close to their graduation and with a similar level of knowledge. Hence, the subjects form a homogeneous group which counteracts the threat of erroneous conclusions. A homogeneous group mitigated the risk that the variation due to the subjects' random heterogeneity is larger than due to the investigated linking variants. Despite the benefit of objective measures, we had to determine the scan-paths (sequence of visited AOIs arranged in chronological order) manually over both artifacts due to a technical constraint of the used eye tracking system (see section 4.3). This manual determination represented a threat to the reliability of our results. We counteracted this threat to validity by including in the manually created scan-paths only the AOI visits that were captured by the eye tracking system. The small sample size of 15 subjects in the baseline experiment represented a further risk that we drew an erroneous conclusion. We minimized this risk by replicating the experiment with 15 new subjects and also analyzing the data of both experiments jointly as if we were dealing with one single experiment. By this means, we could verify our findings and increase their validity. Nevertheless, the total number of 30 subjects (ten per group) is still a limitation of this study. For this reason, further replications of the experiment are necessary to increase the overall validity of the findings.

## 9 Discussion

In the following, we first discuss the findings of the baseline experiment, its repetition, and their joint analysis before presenting planned future work.

### 9.1 Findings of the Experiments

We selected a brain-based IT artifact evaluation for our experimental design to investigate reading behavior by focusing on the brain activity as the mediator (see Figure 4). This brain activity and thus the reading behavior can be described in terms of visual effort and cognitive load.

Our findings show that all three linking variants cause comparable visual effort and share the most frequent sequential pattern of reading the single artifacts separately and successively. Only *integrated links* have an impact on the reading behavior in terms of interrelating both artifacts as an inherent part of its frequent sequential reading patterns.

Our explanation of these findings is based on the *Cognitive Load Theory* by Sweller et al. (2011) which we briefly summarize in the following. The cognitive load determines the required working memory resources of a human brain to process information of given materials. If the cognitive load exceeds the available working memory resources, a human will fail, at least in parts, to process the materials. The total cognitive load imposed by the used materials consists additively of the *intrinsic* and the *extraneous* cognitive load. While the intrinsic cognitive load depends on the nature of the materials in terms of their content difficulty and complexity, extraneous cognitive load depends on the representation and design of the materials. Both single loads and thus the total cognitive load are mainly influenced by the element interactivity of materials. Sweller et al. (2011, p. 58) explain that those "interacting elements are defined as elements that must be processed simultaneously in working memory because they are logically related". The higher the element interactivity the higher is the total cognitive load since more working memory resources are necessary to keep in mind the related information. According to Sweller et al. (2011), a reader needs to process all related information simultaneously in order to understand the overall content. A successive processing of the particular artifacts only enables the understanding of the single materials but not of their interrelationships. Therefore, interrelating all provided materials is an essential part of the reading behavior in order to process and understand them.

In consideration of the *Cognitive Load Theory* (Sweller et al., 2011), our used materials (*fully dressed use case* and natural language requirements) have a high element interactivity since they are strongly associated and represented in a split presentation. This high element interactivity results in a high total cognitive load. This load is increased even further since all subjects were unfamiliar with the materials. Especially, novel information with high element interactivity is likely to impose a high cognitive load.

Corresponding to our experimental design (see section 4.1), we used the same use case and requirements for all subjects to cancel out the effect of different contents. As a consequence, the intrinsic cognitive load is the same for all our subjects since we used the same contents for all three linking variants. Nevertheless, this load is high due to complex interrelationships between the use case and the requirements. Our observations support this assumption because no subject of the group *no linking* interrelated any use case field and corresponding associated requirement on his own.

The extraneous cognitive load needs to be different between the three groups due to the distinct linking representations. One would assume that the visualized links increase the extraneous cognitive load since they need to be processed and understood. However, our results show no significant difference in the reading behavior in terms of visual effort. Hence, the visualized links do not impact the extraneous cognitive load negatively. Instead, the physical integration of links changes the representation of information by supporting the element interactivity and facilitating split-attention. The integration of links externalizes knowledge about interacting elements that must be processed simultaneously in the working memory. The visualized element interactivity frees working memory resources of a reader since fewer elements need to be kept in the working memory. Thus, the links reduce the intrinsic cognitive load by simplifying the identification of interrelationships between the use case fields and associated requirements. A reader can use the freed working memory resources to interrelate both artifacts by processing the related elements simultaneously in order to achieve a good understanding of the overall content. The findings of our experiments support this explanation. *Additional field* and *no linking* lack the explicit externalized knowledge about related use case fields and requirements. In contrast to the *integrated links*, the *additional field* variant is not different from the *no linking* variant. Thus, the knowledge about requirements that are associated with the whole use case is not sufficient for a reader to identify which particular use case fields refer to specific requirements. As a consequence, interrelating the two artifacts is not a trivial task and requires a lot of working memory resources which results in a high total cognitive load. Based on our findings, only *integrated links* achieve additionally available working memory resources due to the externalized knowledge visualized by the physically integrated links. These idle resources were used by a reader to interrelate both artifacts. Our findings support this interpretation since interrelating both artifacts is only part of the frequent sequential patterns of the *integrated links* variants. The joint analysis showed that the efforts to interrelate both artifacts in terms of directed attention switching frequency is only significantly different between the *no linking* and *integrated links* condition with an average increase of 78.1% attention switches in the case of the most detailed linking variant. While we initially inferred these explanations and interpretations only from the findings of the baseline experiment, almost all the findings of the repetition are consistent with the findings of the baseline experiment and thus support these explanations and interpretations as well. Although there was one inconsistency between both experiments, their joint analysis resolved this issue. In this way, the repetition strengthens the validity of our findings. Thus, the repetition helped us to improve the reliability and validity of the explanations and interpretations presented.

The benefit of our results for practice is the insight that the reading behavior can be influenced by the particular linking variant in order to achieve the intended way of interrelating artifacts. As a consequence, it is not trivial to select a use case template. If linking and interrelating artifacts is necessary, *integrated links* are the best option. Although the effort to create such detailed links is higher compared to the other two linking variants, there is a large resulting benefit. An *additional field* does not provide the same effect as *integrated links*. Instead, listing all links in an *additional field* leads to the same reading behavior as *no linking*. We assume that the effort to create and maintain *integrated links* should be comparable to the effort of an *additional field* since the widespread digital maintenance of links is independent of the position of a link. Therefore, the *integrated links* variant is the preferred linking variant since these links are good means to support element interactivity by reducing the intrinsic cognitive load.

All in all, we conclude that the particular linking variant impacts the reading behavior. Even though all three major linking variants cause comparable visual effort and share the most frequent sequential reading pattern, only the *integrated links* caused the intended way of interrelating the two artifacts according to our results.

## 9.2 Future Work

First of all, our findings emphasize the importance of traceability as one key characteristic of good requirements specifications not only as a benefit for later change impact analysis or document maintenance but also for simply reading the document. In consideration of our findings, developers can benefit from the most detailed linking variant by guiding their reading behavior. There is the need for future research to improve support of creating and maintaining *integrated links*. On the one hand, we need to investigate whether use case templates that use the linking variant *no linking* or *additional field* can be adapted to get the benefits of *integrated links*. On the other hand, suitable methods are necessary to encounter the problem of *integrated links* based on identification numbers that are one specific source of risky, dispersed changes of a use case (Basirati et al., 2015). In this way, we can reduce maintenance costs and support readers by interrelating artifacts of a requirements specification more easily.

Our next steps specifically include the following two topics: (a) analysis of understanding and (b) application of *integrated links* across multimedia artifacts.

The topic (a) focuses on the analysis of the readers' understanding. The experiments conducted only investigated the readers' observable reading behavior without examining their concrete understanding of the content. For this purpose, we follow the work of Sanches et al. (2017, 2018) who investigated the readers' understanding of documents by using subjective comprehension questions and objective eye gaze data. Based on the work of Sanches et al. (2017, 2018), we want to adapt the experimental design of our eye tracking study to create further replications of our experiment according to the systematic approach for replication of experiments in software engineering by Gómez et al. (2014). In this way, on the one hand, we increase the sample size to strengthen the conclusion validity of our findings. On the other hand, we collect the necessary information to reuse the analysis approach by Sanches et al. (2017, 2018) that combines subjective ratings in form of Likert scales with objective data, used for an Support Vector Regression (SVR) model.

The topic (b) focuses on our long-term goal of enriching requirements specifications with multimedia artifacts such as videos (Karras et al., 2016a,b, 2017a,c). Especially, in the context of requirements elicitation and validation, the combination of textual artifacts, such as scenarios or use cases, with various multimedia artifacts is an ongoing research topic (Maiden et al., 2004; Rabiser et al., 2006; Seyff et al., 2009, 2010, 2019). In this context, Simmet (2017) developed a prototype for the coevolution of multimedia requirements focusing on creating and maintaining links between textual scenarios and various multimedia artifacts such as videos, audio files, and images. Based on the work of Simmet (2017), we found the need to investigate how links should be created to guide a reader to process interrelated artifacts simultaneously despite their high element interactivity. The experiments conducted lay the foundation for this understanding to guide a reader accordingly.

## 10  Conclusion

The particular linking variant of a use case with its associated requirements has an impact on the reading behavior.

The linking of use cases and requirements is mainly realized by using identification numbers that correspond to the labels of requirements. Besides *no linking*, these labels are either enumerated in an *additional field* or represented as *integrated links* in typical use case fields. Regardless of the applied linking variant, a reader should interrelate a use case and requirements on his own to process and understand both artifacts for themselves and their interrelationships. Created links are intended to support such a reading behavior that includes interrelating both artifacts. However, creating and maintaining links is troublesome since they cause effort and can easily lead to risky, dispersed changes.

In 2017, we performed an eye tracking study to investigate the three previously mentioned linking variants and their impact on the reading behavior in terms of visual effort and intended way of interrelating both artifacts. We repeated this experiment to verify our findings and increase their conclusion validity.

Based on our results, we identified that all three linking variants cause comparable visual effort. Thus, adding links to a use case does not impede its reading. All investigated linking variants also share the most frequent sequential reading pattern. Regardless of the linking variants, all subjects read the use case and the requirements separately and successively. However, we identified a statistically significant difference between the number of directed attention switches by *integrated links* and *no linking*. These results show that only *integrated links* cause more attention switches from a use case to the requirements which represent increased efforts to interrelate both artifacts. The scan-path analysis also showed that the reading behavior of interrelating both artifacts was only part of frequent sequential patterns in the case of *integrated links*.

It is important to emphasize that these findings should not be overgeneralized. Both experiments had an artificial setting that simulated the work of developers who need to process and understand given artifacts of a requirements specification. In contrast to our subjects who were undergraduate and graduate students, more experienced subjects might read the given materials differently. Nevertheless, the reported repetition confirmed almost all results of the baseline experiment. We could clarify the only inconsistent finding between both experiments with the joint analysis.

Our work indicates that all linking variants do not impede the reading of the two artifacts for themselves. However, the specific reading behavior of interrelating both artifacts is only supported by the detailed integration of links. Based on our findings, we recommend preferring the most detailed linking variant *integrated links*.

# References

Ahrens, M., Schneider, K., and Kiesling, S. (2016). How Do We Read Specifications? Experiences from an Eye Tracking Study. In *22nd International Working Conference on Requirements Engineering: Foundation for Software Quality (REFSQ)*, pages 301–317. Springer International Publishing.

Alexander, I. F. and Neil, M. (2005). *Scenarios, Stories, Use Cases: Through the Systems Development Life-cycle*. John Wiley & Sons.

Ali, N., Sharafl, Z., Guéhéneuc, Y.-G., and Antoniol, G. (2012). An Empirical Study on Requirements Traceability Using Eye-Tracking. In *28th IEEE International Conference on Software Maintenance (ICSM)*, pages 191–200. IEEE.

Anda, B., Sjøberg, D. I. K., and Jørgensen, M. (2001). Quality and Understandability of Use Case Models. In *15th European Conference on Object-Oriented Programming (ECOOP)*, pages 402–428. Springer-Verlag.

Araujo, J. and Coutinho, P. (2003). Identifying Aspectual Use Cases Using a Viewpoint-Oriented Requirements Method. In *Early Aspects 2003: Aspect-Oriented Requirements Engineering and Architecture Design*, pages 1–6.

Ayres, J., Flannick, J., Gehrke, J., and Yiu, T. (2002). Sequential Pattern Mining Using a Bitmap Representation. In *8th ACM SIGKDD International Conference on Knowledge Discovery and Data Mining*, pages 429–435.

Basili, V. R., Green, S., Laitenberger, O., Lanubile, F., Shull, F., Sørumgård, S., and Zelkowitz, M. V. (1996). The Empirical Investigation of Perspective-based Reading. *Empirical Software Engineering*, 1(2):133–164.

Basirati, M. R., Femmer, H., Eder, S., Fritzsche, M., and Widera, A. (2015). Understanding Changes in Use Cases: A Case Study. In *23rd IEEE International Requirements Engineering Conference (RE)*, pages 352–361.

Bednarik, R. and Tukiainen, M. (2006). An Eye-Tracking Methodology for Characterizing Program Comprehension Processes. In *Symposium on Eye Tracking Research and Applications (ETRA)*, pages 125–132. ACM.

Berling, T. and Thelin, T. (2004). A Case Study of Reading Techniques in a Software Company. In *International Symposium on Empirical Software Engineering (ISESE04)*, pages 229–238. IEEE.

Binkley, D., Davis, M., Lawrie, D., Maletic, J. I., Morrell, C., and Sharif, B. (2013). The Impact of Identifier Style on Effort and Comprehension. *Empirical Software Engineering*, 18(2):219–276.

Bittner, K. and Spence, I. (2003). *Use Case Modeling*. Addison-Wesley Professional.

Busjahn, T., Schulte, C., and Busjahn, A. (2011). Analysis of Code Reading to Gain More Insight in Program Comprehension. In *11th Koli Calling International Conference on Computing Education Research*, pages 1–9. ACM.

Cockburn, A. (2001). *Writing Effective Use Cases*. Addison-Wesley Reading.

Cohen, J. (2013). Statistical Power Analysis for the Behavioral Sciences.

Coleman, D. (1998). A Use Case Template: Draft for Discussion. *Use Case Template Guidelines*.

El-Attar, M. and Miller, J. (2009). A Subject-Based Empirical Evaluation of SSUCD's Performance in Reducing Inconsistencies in Use Case Models. *Empirical Software Engineering*, 14(5):477–512.

Exadaktylos, F., Espín, A. M., and Brañas-Garza, P. (2013). Experimental Subjects are not Different. *Scientific Reports*, 3.

Fagan, M. (2002). Design and Code Inspections to Reduce Errors in Program Development. *Software Pioneers*, pages 575–607.

Fleisch, W. (1999). Applying Use Cases for the Requirements Validation of Component-Based Real-Time Software. In *Proceedings 2nd IEEE International Symposium on Object-Oriented Real-Time Distributed Computing*, pages 75–84. IEEE.

Fleiss, J. L. (1971). Measuring Nominal Scale Agreement Among Many Raters. *Psychological Bulletin*, 76(5):378–382.

Fricker, S. A., Grau, R., and Zwingli, A. (2015). Requirements Engineering: Best Practice. In *Requirements Engineering for Digital Health*, pages 25–46. Springer International Publishing.

Fusaro, P., Lanubile, F., and Visaggio, G. (1997). A Replicated Experiment to Assess Requirements Inspection Techniques. *Empirical Software Engineering*, 2(1):39–57.

Gómez, O. S., Juristo, N., and Vegas, S. (2014). Understanding Replication of Experiments in Software Engineering: A Classification. *Information and Software Technology*, 56(8):1033–1048.

Gross, A. and Doerr, J. (2012a). What do Software Architects Expect from Requirements Specifications? Results of Initial Explorative Studies. In *1st IEEE International Workshop on the Twin Peaks of Requirements and Architecture*, pages 41–45.

Gross, A. and Doerr, J. (2012b). What You Need is What You Get! The Vision of View-based Requirements Specifications. In *20th IEEE International Requirements Engineering Conference (RE)*, pages 171–180.

Halling, M., Biffl, S., Grechenig, T., and Kohle, M. (2001). Using Reading Techniques to Focus Inspection Performance. In *27th EUROMICRO Conference*, pages 248–257.

Harwood, R. J. (1997). Use Case Formats: Requirements, Analysis, and Design. *Journal of Object-Oriented Programming*, 9(8).

Haumer, P. (2004). *Use Case-Based Software Development*. IBM Rationale Rose.

Hejmady, P. and Narayanan, N. H. (2012). Visual Attention Patterns During Program Debugging with an IDE. In *Symposium on Eye Tracking Research and Applications (ETRA)*, pages 197–200. ACM.

Höst, M., Regnell, B., and Wohlin, C. (2000). Using Students as Subjects – A Comparative Study of Students and Professionals in Lead-Time Impact Assessment. *Empirical Software Engineering*, 5(3).

Insfrán, E., Pastor, O., and Wieringa, R. (2002). Requirements Engineering-Based Conceptual Modelling. *Requirements Engineering*, 7(2):61–72.

Jaaksi, A. (1998). Our Cases with Use Cases.

Jacobson, I. (1993). *Object-Oriented Software Engineering: A Use Case Driven Approach*. Pearson Education India.

Karras, O. (2021). Eye Tracking Experiments Data Set - Linking Use Cases and Associated Requirements: On the Impact of Linking Variants on Reading Behavior. Zenodo.

Karras, O., Hamadeh, A., and Schneider, K. (2017a). Enriching Requirements Specifications with Videos – The Use of Videos to Support Requirements Communication. *GI Softwaretechnik-Trends*, 38(1).

Karras, O., Kiesling, S., and Schneider, K. (2016a). Supporting Requirements Elicitation by Tool-Supported Video Analysis. In *2016 IEEE 24th International Requirements Engineering Conference (RE)*. IEEE.

Karras, O., Klünder, J., and Schneider, K. (2016b). Enrichment of Requirements Specifications with Videos: Enhancing the Comprehensibility of Textual Requirements. In *Proceedings of Videos in Digital Libraries – What's in it for Libraries, Scientists, and Publishers? (TPDL)*. Zenodo.

Karras, O., Klünder, J., and Schneider, K. (2017b). Is Task Board Customization Beneficial? In *International Conference on Product-Focused Software Process Improvement*, pages 3–18. Springer.

Karras, O., Risch, A., and Schneider, K. (2018). Interrelating Use Cases and Associated Requirements by Links: An Eye Tracking Study on the Impact of Different Linking Variants on the Reading Behavior. In *Proceedings of the 22nd International Conference on Evaluation and Assessment in Software Engineering 2018*, EASE'18, New York, NY, USA. ACM.

Karras, O., Unger-Windeler, C., Glauer, L., and Schneider, K. (2017c). Video as a By-Product of Digital Prototyping: Capturing the Dynamic Aspect of Interaction. In *2017 IEEE 25th International Requirements Engineering Conference Workshops (REW)*. IEEE.

Kettenis, J. (2007). Getting Started with Use Case Modeling: White Paper. *Oracle Corporation*.

Kruchten, P. (2004). *The Rational Unified Process: An Introduction*. Addison-Wesley Professional.

Kujala, S., Kauppinen, M., and Rekola, S. (2001). Bridging the Gap Between User Needs and User Requirements. pages 45–50.

Kulak, D. and Guiney, E. (2012). *Use Cases: Requirements in Context*. Addison-Wesley.

Landis, J. R. and Koch, G. G. (1977). The Measurement of Observer Agreement for Categorical Data. *Biometrics*, pages 159–174.

Leite, J. C. S. d. P., Hadad, G. D., Doorn, J. H., and Kaplan, G. N. (2000). A Scenario Construction Process. *Requirements Engineering*, 5(1):38–61.

Liu, D., Subramaniam, K., Far, B. H., and Eberlein, A. (2003). Automating Transition from Use Cases to Class Model. In *Canadian Conference on Electrical and Computer Engineering. Toward a Caring and Humane Technology*, volume 2, pages 831–834. IEEE.

Maiden, N., Seyff, N., and Grunbacher, P. (2004). The Mobile Scenario Presenter: Integrating Contextual Inquiry and Structured Walkthroughs. In *13th IEEE International Workshops on Enabling Technologies: Infrastructure for Collaborative Enterprises*. IEEE.

Mattingly, L. and Rao, H. (1998). Writing Effective Use Cases and Introducing Collaboration Cases. *Journal of Object-Oriented Programming*, 11(6).

Mich, L., Franch, M., and Novi Inverardi, P. (2004). Market Research for Requirements Analysis Using Linguistic Tools. *Requirements Engineering*, 9(1):40–56.

Miller, J., Wood, M., and Roper, M. (1998). Further Experiences with Scenarios and Checklists. *Empirical Software Engineering*, 3(1):37–64.

Misbhauddin, M. and Alshayeb, M. (2015). Extending the UML Use Case Metamodel with Behavioral Information to Facilitate Model Analysis and Interchange. *Software & Systems Modeling*, 14(2):813–838.

Öuergaard, G. K. P. (2005). Use Gases-Patterns and Blueprints.

Paech, B. and Kohler, K. (2004). Task-Driven Requirements in Object-Oriented Development. *Kluwer International Series in Engineering and Computer Science*, pages 45–68.

Porras, G. C. and Guéhéneuc, Y.-G. (2010). An Empirical Study on the Efficiency of Different Design pattern Representations in UML Class Diagrams. *Empirical Software Engineering*, 15(5):493–522.

Porter, A. A. and Votta, L. G. (1998). Comparing Detection Methods for Software Requirements Inspections: A Replication Using Professional Subjects. *Empirical software engineering*, 3(4):355–379.

Porter, A. A., Votta, L. G., and Basili, V. R. (1995). Comparing Detection Methods for Software Requirements Inspections: A Replicated Experiment. *IEEE Transactions on Software Engineering*, 21(6):563–575.

Rabiser, R., Seyff, N., Grunbacher, P., and Maiden, N. (2006). Capturing Multimedia Requirements Descriptions with Mobile RE Tools. In *2006 First International Workshop on Multimedia Requirements Engineering*. IEEE.

Riedl, R. and Léger, P.-M. (2016). *Fundamentals of NeuroIS*. Springer.

Robertson, S. and Robertson, J. (2012). *Mastering the Requirements Process: Getting Requirements Right*. Addison-Wesley.

Romero, P., Cox, R., du Boulay, B., and Lutz, R. (2002). Visual Attention and Representation Switching During Java Program Debugging: A Study Using the Restricted Focus Viewer. *Diagrammatic Representation and Inference*, pages 323–326.

Romero, P., du Boulay, B., Cox, R., and Lutz, R. (2003). Java Debugging Strategies in Multi-Representational Environments. In *15th Annual Workshop of the Psychology of Programming Interest Group (PPIG)*, pages 421–434.

Runeson, P. (2003). Using Students as Experiment Subjects – An Analysis on Graduate and Freshmen Student Data. In *Proceedings of the 7th International Conference on Empirical Assessment in Software Engineering*.

Sanches, C. L., Augereau, O., and Kise, K. (2017). Using the Eye Gaze to Predict Document Reading Subjective Understanding. In *14th IAPR International Conference on Document Analysis and Recognition*, volume 8. IEEE.

Sanches, C. L., Augereau, O., and Kise, K. (2018). Estimation of Reading Subjective Understanding based on Eye Gaze Analysis. *PloS one*, 13(10).

Sandahl, K., Blomkvist, O., Karlsson, J., Krysander, C., Lindvall, M., and Ohlsson, N. (1998). An Extended Replication of an Experiment for Assessing Methods for Software Requirements Inspections. *Empirical Software Engineering*, 3(4):327–354.

Santos, M., Gralha, C., Goulão, M., Araújo, J., Moreira, A., and Cambeiro, J. (2016). What is the Impact of Bad Layout in the Understandability of Social Goal Models? In *24th IEEE International Requirements Engineering Conference (RE)*, pages 206–215.

Schneider, G. and Winters, J. P. (1998). *Applying Use Cases: A Practical Guide*. Addison Wesley.

Seyff, N., Graf, F., and Maiden, N. (2010). Using Mobile RE Tools to Give End-Users their own Voice. In *18th IEEE International Requirements Engineering Conference*. IEEE.

Seyff, N., Maiden, N., Karlsen, K., Lockerbie, J., Grünbacher, P., Graf, F., and Ncube, C. (2009). Exploring how to use scenarios to discover requirements. *Requirements Engineering*, 14(2).

Seyff, N., Vierhauser, M., Schneider, M., and Cleland-Huang, J. (2019). Towards the Next Generation of Scenario Walkthrough Tools – A Research Preview. In *International Working Conference on Requirements Engineering: Foundation for Software Quality*. Springer.

Sharafi, Z., Marchetto, A., Susi, A., Antoniol, G., and Guéhéneuc, Y.-G. (2013). An Empirical Study on the Efficiency of Graphical vs. Textual Representations in Requirements Comprehension. In *21st International Conference on Program Comprehension (ICPC)*, pages 33–42.

Sharafi, Z., Soh, Z., and Guéhéneuc, Y.-G. (2015). A Systematic Literature Review on the Usage of Eye-Tracking in Software Engineering. *Information and Software Technology*, 67:79–107.

Sharafi, Z., Soh, Z., Guéhéneuc, Y.-G., and Antoniol, G. (2012). Women and Men – Different but Equal: On the Impact of Identifier Style on Source Code Reading. In *20st International Conference on Program Comprehension (ICPC)*, pages 27–36.

Sharif, B. and Maletic, J. I. (2010). An Eye Tracking Study on the Effects of Layout in Understanding the Role of Design Patterns. In *IEEE International Conference on Software Maintenance (ICSM)*, pages 1–10.

Shull, F. (2002). Software Reading Techniques. *Encyclopedia of Software Engineering*.

Simmet, L. (2017). Coevolution of Multimedia Requirements. Master thesis, Leibniz Universität Hannover.

Sjøberg, D. I. K., Anda, B., Arisholm, E., Dybå, T., Jørgensen, M., Karahasanovic, A., Koren, E. F., and Vokác, M. (2002). Conducting Realistic Experiments in Software Engineering. In *Proceedings International Symposium on Empirical Software Engineering*. IEEE.

Somé, S. S. (2006). Supporting use case based requirements engineering. *Information and Software Technology*, 48(1):43–58.

Sweller, J., Ayres, P., and Kalyuga, S. (2011). *Cognitive Load Theory*, volume 1.

Thelin, T., Runeson, P., and Regnell, B. (2001). Usage-based Reading – An Experiment to Guide Reviewers with Use Cases. *Information and Software Technology*, 43(15):925–938.

Thelin, T., Runeson, P., and Wohlin, C. (2003). An Experimental Comparison of Usage-based and Checklist-based Reading. *IEEE Transactions on Software Engineering*, 29(8):687–704.

Tiwari, S. and Gupta, A. (2015). A Systematic Literature Review of Use Case Specifications Research. *Information and Software Technology*, 67:128–158.

Tiwari, S., Rathore, S. S., Gupta, S., Gogate, V., and Gupta, A. (2012). Analysis of Use Case Requirements using SFTA and SFMEA Techniques. In *17th International Conference on Engineering of Complex Computer Systems*, pages 29–38. IEEE.

Toro, A. D., Bernárdez Jiménez, B., Ruiz Cortés, A., and Toro Bonilla, M. (1999). A Requirements Elicitation Approach based in Templates and Patterns.

Travassos, G., Shull, F., Fredericks, M., and Basili, V. R. (1999). Detecting Defects in Object-Oriented Designs: Using Reading Techniques to Increase Software Quality. In *14th ACM SIGPLAN Conference on Object-Oriented Programming, Systems, Languages, and Applications (OOPSLA)*, pages 47–56.

Uwano, H., Nakamura, M., Monden, A., and Matsumoto, K.-i. (2006). Analyzing Individual Performance of Source Code Review Using Reviewers' Eye Movement. In *Symposium on Eye Tracking Research and Applications (ETRA)*, pages 133–140. ACM.

Wiegers, K. and Beatty, J. (2013). Software requirements.

Wiegers, K. E. (1999). Automating Requirements Management. *Software Development*, 7(7):1–5.

Wohlin, C., Runeson, P., Höst, M., Ohlsson, M. C., Regnell, B., and Wesslén, A. (2012). *Experimentation in Software Engineering*. Springer.

Yue, T., Briand, L. C., and Labiche, Y. (2013). Facilitating the Transition from Use Case Models to Analysis Models: Approach and Experiments. *ACM Transactions on Software Engineering and Methodology*, 22(1):1–38.

Yusuf, S., Kagdi, H., and Maletic, J. I. (2007). Assessing the Comprehension of UML Class Diagrams via Eye Tracking. In *15th IEEE International Conference on Program Comprehension (ICPC)*, pages 113–122.

Zhou, J., Lu, Y., Lundqvist, K., Lönn, H., Karlsson, D., and Liwång, B. (2014). Towards Feature-Oriented Requirements Validation for Automotive Systems. In *22nd IEEE International Requirements Engineering Conference (RE)*, pages 428–436.

Zhu, Y.-M. (2016). *Software Reading Techniques*. Springer.